\documentclass[a4paper,fleqn]{cas-dc}

\usepackage{soul}
\usepackage{color, xcolor} 
\usepackage[numbers]{natbib}
\usepackage{tabularx}
\usepackage{graphicx}
\usepackage{caption}
\usepackage{url}
\usepackage[switch]{lineno}

\begin{document}

\shorttitle{AI-Generated Content in Landscape Architecture: A Survey} 


\title [mode = title]{AI-Generated Content in Landscape Architecture: A Survey}

\author[1]{Yue Xing}
\ead{xingyue9812@gmail.com}
\address[1]{School of Art and Design, Shaoguan University, Shaoguan 512005, China} 

\author[2]{Wensheng Gan}
\cortext[cor1]{Corresponding author}
\ead{wsgan001@gmail.com}
\address[2]{College of Cyber Security, Jinan University, Guangzhou 510632, China} 
\cormark[1]

\author[1]{Qidi Chen}
\ead{dylanc0829@gmail.com}

\author[3]{Philip S. Yu}
\ead{psyu@uic.edu}
\address[3]{Department of Computer Science, University of Illinois Chicago, Chicago 60607, USA}

\begin{abstract}
  Landscape design is a complex process that requires designers to engage in intricate planning, analysis, and decision-making. This process involves the integration and reconstruction of science, art, and technology. Traditional landscape design methods are shaped by factors such as the designer's knowledge, time constraints, ecological climate, available resources, and environmental considerations. These methods often rely on the designer's personal experience and subjective aesthetics, with design standards rooted in subjective perception. As a result, they lack scientific and objective evaluation criteria and systematic design processes. Data-driven artificial intelligence (AI) technology provides an objective and rational design process. With the rapid development of different AI technologies, AI-generated content (AIGC) has permeated various aspects of landscape design at an unprecedented speed, serving as an innovative design tool. This article aims to explore the applications and opportunities of AIGC in landscape design. AIGC can support landscape design in areas such as site research and analysis, design concepts and scheme generation, parametric design optimization, plant selection and visual simulation, construction management, and process optimization. However, AIGC also faces challenges in landscape design, including data quality and reliability, design expertise and judgment, technical challenges and limitations, site characteristics and sustainability, user needs and participation, the balance between technology and creativity, ethics, and social impact. Finally, this article provides a detailed outlook on the future development trends and prospects of AIGC in landscape design. Through in-depth research and exploration in this review, readers can gain a better understanding of the relevant applications, potential opportunities, and key challenges of AIGC in landscape design.
\end{abstract}


\maketitle

\section{Introduction}  \label{sec:Introduction}

Landscape architecture (LA) \cite{francis2001case,newton1971design} is the art and science of creating a harmonious relationship between humans and nature and providing spaces for human interaction. It combines creativity with technical expertise and involves a complex design process that requires extensive planning, analysis, and decision-making by designers. It is a process of integrating and reconstructing science, art, and technology \cite{eckbo1974art}. The planning and design process of LA involves comprehensive analysis, optimization, and reconstruction of objective elements such as geographical environment, site, transportation, architecture, and plants \cite{marsh2005landscape}. It also requires the use of different artistic forms to express the specific spirit and cultural value of a place. In the context of LA, both humans and nature are essential elements of the landscape \cite{tress2001bridging}. Specifically, "landscape" is the manifestation of human perception and practice of the natural environment, and "humans" are one of the indispensable key elements in the landscape. Therefore, LA planning and design should be human-centered \cite{buchanan2001human}. With the rapid development of urbanization, LA design should be based on regional characteristics and integrate humanistic, natural, and ecological elements into the landscape design. This approach creates works that are well suited to the environment, harmonizing with their surroundings and better serving the city. Currently, traditional LA design methods \cite{filor1994nature} are influenced by various factors such as designers' knowledge, time constraints, climate change, ecological degradation, resource and environmental issues, human history, social equity, and cultural constraints. These design methods frequently rely on designers' personal experience and aesthetics, which can subjectively limit design outcomes. Due to the lack of scientific and objective evaluation standards and processes, design standards remain at a subjective and sensory level, leaning towards art rather than science, design rather than planning, and description rather than evaluation. However, data-driven artificial intelligence (AI) \cite{nilsson2014principles} provides an objective and rational way to optimize the design process.

With the development of AI and exploration in various disciplines, landscape architecture, as an interdisciplinary field \cite{chen2024Origin}, has begun to focus on and apply AI technology to promote sustainable design and innovation \cite{xing2024artificial}. AI involves various intelligent technologies such as machine learning (ML) \cite{alpaydin2020introduction,jordan2015machine}, data mining \cite{gan2023anomaly,gan2017data,wang2023ai}, natural language processing (NLP) \cite{collobert2011natural}, and image recognition \cite{he2016deep}. These AI technologies are based on rich data and form the foundation of AI. Among them, AI-generated content (AIGC) \cite{wu2023ai} provides a new technological path for the development of AI. It has received widespread attention and exploration in LA, providing landscape architects with new platform tools and design methods. AI can not only provide assistance at the tool and method level in the design field but also inspire the fundamental level of thinking and methodology for the discipline \cite{Gong2024Chinese}. AIGC has significant advantages in empowering LA design, generating construction drawings \cite{rahbar2022architectural}, and creating interactive visual user experiences \cite{portman2015go}. AIGC improves design efficiency, optimizes design solutions, and automates key design and analytical decision-making processes in LA through the use of ML \cite{balasubramanian2022substituting}, big data analysis \cite{gan2019survey,sagiroglu2013big}, and algorithm optimization. This includes planning and design \cite{peng2021intelligent}, site analysis \cite{zhang2019joint}, plant configuration \cite{liu2022application}, design generation \cite{chang2021building}, optimization, construction and management \cite{pan2021roles}, etc., thereby improving design efficiency and quality. The seamless integration of ecology, art, culture, and society is the goal of modern LA design \cite{xing2024artificial}. The application opportunities of AIGC in the field of LA include the following aspects:

1) Investigation phase of LA design projects. "In any given environmental context, natural, cultural, and aesthetic elements have historical inevitability. Designers must fully understand them before they can decide what should happen in this environment." - Lawrence Halprin. Firstly, AI can perform environmental analysis through big data analysis \cite{gan2021explainable,stupariu2022machine}, gathering relevant background information about the region, including cultural customs, geography, climate, and soil to understand the characteristics and constraints of the design site. Then, AIGC can quickly collect and analyze a large amount of project-related data through data mining \cite{fournier2022pattern,gan2017data} and data analysis \cite{davenport2018analytics}, extract valuable information, and provide comprehensive project background information and environmental analysis reports. For example, in landscape design, when designing urban rail transit hub plazas, emphasis is placed on analyzing location, transportation, and functional land to understand the project's design environment. This includes the site's location, regional environmental characteristics, site scale characteristics, distribution of major transportation routes, public transportation distribution, natural landscape features such as hydrology and soil, as well as functional land uses such as commercial, educational, and residential areas.

2) During the concept and schematic design phase of LA, AIGC can automatically generate multiple design concepts and schemes based on the design task using deep learning (DL) \cite{senem2023using} and generative adversarial networks (GANs) \cite{chen2023generative}. Designers can use AIGC technology to quickly generate and evaluate different design options, speeding up the design process. It reduces the exploration time in the sketch design phase and the modeling time for repeated modifications. The interactive and dynamic visualizations generated by AIGC allow clients to intuitively understand the design intent. Combined with virtual simulation technology \cite{jo2022perception}, AIGC enhances the visualization and persuasiveness of design proposals and reduces communication costs. For example, designers can provide AI with textual information such as landscape style, location, and functionality. Then, AIGC technology generates designs based on the provided input. Designers can provide feedback on the generated design options, and AI generates new optimized design options based on the feedback. This iterative process continues until a satisfactory final design outcome is achieved.

3) During the construction drawing design phase of LA, AIGC can provide automated design tools to assist designers in quickly generating construction drawings and detailed plans \cite{weber2022automated}. Through deep learning \cite{akinosho2020deep,elyan2020deep}, AIGC can identify and optimize the design elements of construction drawings, improving their accuracy and efficiency. For example, during this phase, landscape information modeling (LIM) \cite{abdirad2015advancing,ahmad2012need} integrates the design, structure, and construction information from different professional departments, construction units, design units, and supervisory units involved in the landscape design and construction process. This helps resolve conflicts and better expresses the design intent, bridging the relationships between different disciplines and providing a visual representation of the design.

4) Construction simulation and optimization. AIGC can simulate construction progress, material consumption, and equipment configuration \cite{baduge2022artificial}, providing optimization decisions for construction units. It improves the accuracy of budgeting during the design phase, allowing for detailed cost estimation and reducing the need for extensive adjustments later on, thus reducing construction costs and improving efficiency. Additionally, AIGC can generate virtual construction scenarios to help designers and owners anticipate potential issues and make necessary preparations in advance. Based on various technologies such as building information modeling (BIM) \cite{routledge2016bim,zabin2022applications}, the continuous development and application of AI have introduced more AIGC techniques into LA projects, enabling intelligent construction and management while improving project efficiency and quality \cite{pan2023integrating}.

However, AIGC faces several challenges in LA design. 1) There are issues related to data quality and security \cite{liang2022advances}. AIGC applications rely on a large amount of environmental data and design samples. Ensuring the accuracy, completeness, and privacy of this data is a key problem that needs to be addressed. 2) The professional expertise and judgment of human designers play a crucial role throughout the landscape planning and design process. It requires the full manifestation of their professional spirit and creative aesthetics \cite{amabile2020creativity}. 3) There are technical challenges and limitations due to algorithm biases and uncertainties \cite{an2021determining}. The content generated by AIGC may be influenced by the limitations of training data and algorithms, resulting in certain biases and uncertainties. 4) There is a need to generate designs that match personalized user requirements \cite{fan2006personalization}, necessitating adjustments and customization by designers. 5) Considerations of site characteristics and sustainability are crucial \cite{wu2013landscape}. 6) Balancing technology and art is important, maintaining technological advancement while preserving the independent style and creativity of designers. 7) There are ethical and legal risks associated with the widespread application of AIGC, such as copyright and liability issues \cite{guo2023aigc}. The relevant regulations and ethical guidelines need to be established. In summary, AIGC is rapidly penetrating various aspects of LA design, bringing new opportunities to the field. Practitioners in the LA field should actively embrace AIGC technology and leverage its advantages in fostering creativity, environmental analysis, design development, and optimization. However, they should also acknowledge and address the existing problems and challenges by improving standards, strengthening regulatory mechanisms, and ensuring the safety, compliance, and controllability of AIGC applications. This will make AIGC a powerful driving force for innovative development in LA. 

This review aims to explore the prospects and challenges of AIGC in LA, providing ideas and suggestions to promote innovative development in this field. It hopes to provide a comprehensive perspective for LA practitioners and AI professionals to understand the application and potential of AIGC in LA design, as well as the related technologies, challenges, and prospects. This will contribute to the promotion of innovative development in LA design, as well as create sustainable garden landscapes. The main contributions of this paper are as follows:

\begin{itemize}
    \item  We introduce the main limitations and challenges faced by LA and discuss the inspirations brought by the ongoing technological transformation to design methodologies, mainly manifested in design creativity generation, scene simulation and visualization, plant management optimization, operation and maintenance intelligence, and landscape design optimization (Section  \ref{sec:status}).
    
    \item We specifically discuss the key technologies of AIGC, its applications and potential in LA, and further explore the critical advancements of AIGC in the field of LA (Section  \ref{sec:concepts}).
    
    \item We discuss the specific applications of AIGC in various stages and domains of LA, providing support for site investigation and analysis, design concepts and scheme generation, parametric design optimization, plant configuration and simulation, construction management, and optimization, etc. (Section  \ref{sec:applications}).
    
    \item  We highlight the potential challenges and trends of AIGC in LA. These challenges include data quality and reliability, design professionalism and capabilities, technical challenges and limitations, site characteristics and sustainability, user requirements and involvement, etc. The development trends include multidisciplinary integration, comprehensive coverage of technology, data support and algorithm optimization, close integration with other technologies, and the imperfection of ethics and regulations (Section  \ref{sec:trends}).
\end{itemize}

\begin{figure}[ht]
    \centering
    \includegraphics[width=9cm]{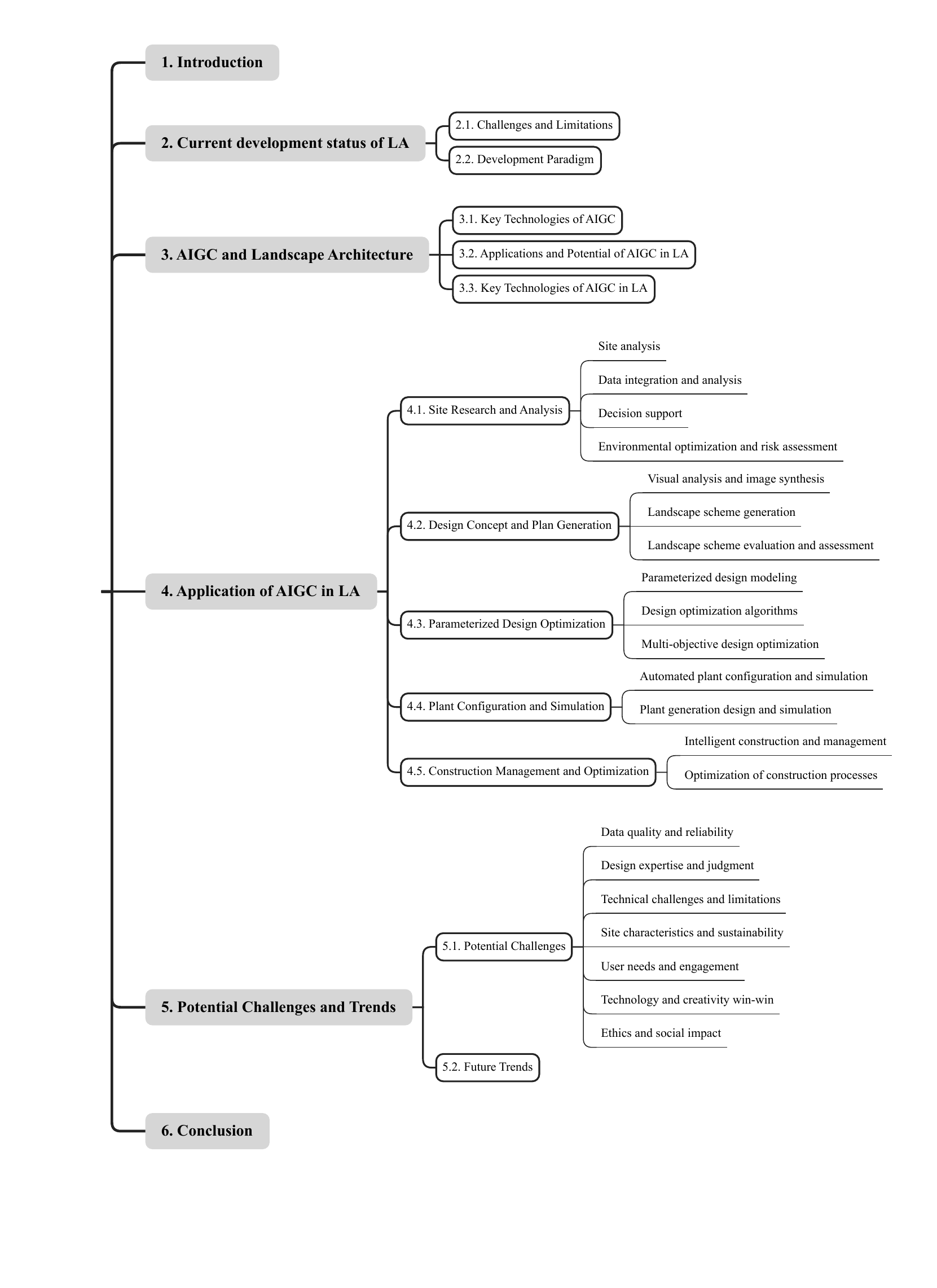}
    \caption{The outline of our overview.}
    \label{fig:outline}
\end{figure}

Some important terms of acronyms in this paper are listed in Table \ref{table:acronyms}.

\begin{table}[ht]
    \centering
    \caption{Important terms of acronyms.}
    \label{table:acronyms}
    \begin{tabular}{|m{1.5cm}<{\raggedright}|m{5.5cm}<{\raggedright}|}
        \hline
        \multicolumn{1}{|c|}{\textbf{Acronym}} & \multicolumn{1}{c|}{\textbf{Full Form}} \\  
        \hline
        AI & Artificial Intelligence \\
        \hline
        AIGC & Artificial Intelligence Generated Content \\
        \hline
        AR & Augmented Reality \\
        \hline
        BERT & Bidirectional Encoder Representations from Transformer \\
        \hline
        BIM & Building Information Modeling \\
        \hline
        CAD & Computer-Aided Design \\
        \hline 
        CNN & Convolutional Neural Networks \\
        \hline
        CPRM & Construction Project Risk Management \\
        \hline
        CV &  Computer Vision \\
        \hline
        DL & Deep Learning \\
        \hline
        DT & Digital Twin \\
        \hline
        GA & Genetic Algorithms \\
        \hline
        GAN & Generative Adversarial Network \\
        \hline
        GNN & Graph Neural Network \\
        \hline
        GIS & Geographic Information System \\
        \hline
        GPT & Generative Pre-trained Transformer \\
        \hline
        GSV & Google Street View \\
        \hline
        I2I & Image-to-Image \\
        \hline
        ISM & Image Style Migration \\
        \hline
        LA & Landscape Architecture \\
        \hline
        LIM & Landscape Information Modeling \\
        \hline
        LLMs & Large Language Models \\
        \hline
        LSTM & Long short-term memory \\
        \hline
        LUCC & Land Use and Land Cover Change \\
        \hline
        ML & Machine Learning \\
        \hline
        NLP & Natural Language Processing \\
        \hline
        PSO & Particle Swarm Optimization \\
        \hline
        RNN & Recurrent Neural Network \\
        \hline
        SA & Simulated Annealing \\
        \hline
        TLS & Terrestrial Laser Scanning \\
        \hline
        TSV & Tencent Street View \\
        \hline
        VAE & Variational Auto-encoder \\
        \hline
        ViT & Vision Transformer \\
        \hline
        VR & Virtual Reality \\
        \hline   
        \end{tabular}
\end{table}

\section{Current development status of LA} \label{sec:status}
\subsection{Challenges and Limitations}

Landscape architecture (LA) requires more innovation, effective management, and analysis of the collected data and information to provide attractive and powerful landscape design solutions that support decision-making and design. With the acceleration of digital transformation, LA needs to adopt digital tools and technologies such as building information modeling (BIM) \cite{routledge2016bim,zabin2022applications}, landscape information modeling (LIM) \cite{abdirad2015advancing,ahmad2012need}, and AIGC to improve efficiency, accuracy, and visualization capabilities \cite{bryan2003physical}. Unified standards and specifications are beneficial for project management, better coordination among relevant departments and roles, and ensuring smooth project progress. LA needs to ensure economic feasibility \cite{budiharta2016enhancing} to achieve cost-effectiveness and return on investment. With population growth and rapid urbanization, LA needs to focus more on sustainability \cite{zhou2019sustainable} to address challenges such as limited water resources and climate change, while also considering environmental and ecological balance to reduce ecosystem destruction. LA addresses issues related to public participation, laws, and policies \cite{buchecker2003participatory,conrad2011research}, and improves communication and coordination with the public to meet the needs of various interest groups. In summary, LA faces various challenges and issues that require proactive measures and the latest technological means to ensure sustainable, innovative, and efficient landscape design. Specifically, the LA design process is mainly limited by subjective experience, design cycles, and regional characteristics:

\textbf{1) Design methods rely on subjective human experience.} It is primarily constrained by the aesthetic concepts and values of designers or clients, lacking systematic quantitative analysis \cite{bryman2006integrating} and support of scientific data \cite{wilkinson2016fair}, resulting in design outcomes biased towards subjectivity \cite{lu2011subjectivity}. In traditional LA design processes, data collection relies entirely on manual work, making it difficult to avoid subjective design decisions, especially when dealing with large and medium-scale environments where limitations are more prominent. Due to the lack of scientific and objective analysis and evaluation criteria and processes, design tends to focus on the subjective and sensory aspects of human experience \cite{lenzholzer2013research}. Note that data-driven AI technology can provide new possibilities for LA design by optimizing the design process objectively and rationally. For example, in traditional design, a designer's personal experience, aesthetics, and preferences may influence design proposals, leading to a lack of holistic, objective data analysis and evaluation.

\textbf{2) Design proposal cycles are often lengthy}. Traditional LA design processes require a significant investment of manpower and time. Tasks such as hand-drawn sketches, site surveys, and plant research consume substantial time and effort, limiting work efficiency and the creativity of designers. For example, traditional methods \cite{jones1992design} have limitations in dealing with complex spatial relationships, landscape environments, and multi-objective optimization. Designers are required to manually explore multiple design options, while this consumes time and resources. For instance, when designing urban squares \cite{matsuoka2008people}, designers will analyze factors such as pedestrian flow, sightlines, and functional requirements, and achieve balance and optimization. Additionally, cost budgeting accuracy is often insufficient, leading to repeated adjustments of design proposals. For example, inadequate cost accuracy during the conceptual phase of a project results in cost overruns and requires repeated adjustments. AI systems, on the other hand, can achieve cost matching during the conceptual design phase and provide detailed cost breakdowns for the generated proposals \cite{elmousalami2020artificial}.

\textbf{3) Influence of regional characteristics}. Design processes may face restrictions related to land use, environmental protection requirements, and regulations \cite{terra2014land}. It is necessary to find a balance in the design and meet the relevant regulations and requirements. For example, when designing urban parks, designers consider urban planning and land use regulations to ensure that design proposals comply with applicable laws and regulations. LA should embody local characteristics, basing designs on regional features such as landforms, geology, topography, and climate, while integrating human, natural, and ecological elements. In the LA design process, these natural conditions and regional factors \cite{philokyprou2017environmentally} are the prerequisites for landscape planning and landscape design.

\textbf{4) Limited public participation}. Stakeholders in a landscape design project include regulatory groups (such as government planning departments and local governments), design and implementation groups (such as urban planners, designers, and engineers), and the public group (including experts, residents, end-users, and community committees) \cite{quan2019artificial}. The emergence of shared digital models (such as city information modeling and BIM) has changed the way designers, experts, and owners collaborate but may hinder the involvement of most experts and community residents with professional or local knowledge in shaping key decisions at the early design stage \cite{zhao2018performance}.

Therefore, it is necessary to deeply understand the limitations of traditional LA design methods and expect AIGC to leverage its potential and advantages in addressing these challenges. In the following sections, we will explore in-depth how AIGC can tackle these challenges and bring opportunities for innovation and improvement in LA design.

\subsection{Development Paradigm}

LA design should consider various environmental elements \cite{booth1989basic}, including urban climate conditions, precipitation, wind direction, and resident needs, such as leisure areas, green vegetation, and water features. Notice that different types of LA projects have their unique requirements, such as urban squares, courtyards, scenic nature reserves, etc., which require personalized solutions based on site characteristics and design goals. For example, when designing trail systems in natural reserves \cite{marion2023trail}, designers consider ecosystem characteristics, conservation requirements for flora and fauna species, as well as visitor safety and experience. The complexity and diversity of LA design stem from various aspects: the diversity of natural environments, design goals, functional uses, cultural backgrounds and historical heritage, social demands, and species diversity. LA faces challenges of diversity and complexity \cite{parrott2012future}, and designers should consider various factors while achieving coordination and unity of design goals to create sufficient urban space. In the AI era, LA design is undergoing profound changes, mainly in the following aspects, including but not limited to:

\textbf{1) Design creativity generation}: Traditional LA design mainly relies on the designer's inspiration and on-site research. The AI models can learn from a large number of landscape design cases, discover hidden creative patterns and landscape aesthetic rules, and provide designers with new ways to generate creativity \cite{doshi2024generative}. Designers can collaborate with AI systems to stimulate richer and more diverse landscape ideas \cite{wu2021ai}.

\textbf{2) Scene simulation and visualization}: Traditional LA modeling is time-consuming and labor-intensive. AI technology empowers LA design with powerful virtual simulation and visualization capabilities \cite{ervin2001digital,orland2001considering}. Designers can use AIGC to generate three-dimensional scene models in minutes, simulate different environmental conditions, vegetation combinations, pedestrian flows, etc., and predict the overall effects of LA, conveying visual and tangible graphic language. AI can transform design proposals into realistic dynamic renderings, allowing clients and users to intuitively understand the design intent. This virtual simulation technology \cite{lovett2015using} enhances the visualization level and attractiveness of design proposals.

\textbf{3) Optimization of plant management}: In landscape design, plant configuration is a crucial aspect. AI technology can combine a large amount of plant growth data to provide designers with intelligent plant selection and configuration solutions \cite{makowski2019synthetic}. It can recommend the optimal plant configuration based on factors such as environmental conditions and aesthetic requirements, and predict plant growth trends \cite{muhar2001three} to help designers optimize the greening layout of LA \cite{feng2022optimal}. This intelligent plant management capability contributes to the practicality and ecological aspects of LA design.

\textbf{4) Intelligent operation and maintenance management}: In the post-construction and utilization phase of LA, AI technology also plays an important role. It can automatically regulate the lighting, irrigation, drainage, and other systems of LA by monitoring environmental data \cite{bwambale2022smart}, such as meteorological conditions and pedestrian flow, optimizing the daily maintenance of the park. Additionally, AI can analyze the dynamic operation status of the park and provide decision support for subsequent maintenance and upkeep. Data-driven intelligent operation and maintenance management \cite{sirvio2015intelligent} significantly improves operational efficiency and user experience within the park.

\textbf{5) Landscape design optimization}: AI technology can also analyze \cite{stupariu2022machine} and optimize \cite{caldas2002design} existing landscape designs. It can collect user feedback and combine professional landscape aesthetic theories to identify issues within the park and provide targeted improvement suggestions to designers. This data-driven landscape evaluation and optimization capability  \cite{jahani2020forest,yu2022optimization} enhances the scientific nature of LA design. Real-time rendering capabilities greatly enhance the efficiency of designers, allowing them to iterate and optimize design proposals quickly, bridging the gap from conceptual design to design expression.

\section{AIGC and Landscape Architecture} \label{sec:concepts}
\subsection{Key Technologies of AIGC}

AI-generated content (AIGC) \cite{wu2023ai} is an important component of AI-related applications. Specifically, AIGC utilizes technologies such as generative adversarial networks (GANs) \cite{goodfellow2014generative}, pre-trained models \cite{zeng2023distributed}, and literature indexing to generate new relevant content, such as news, articles, music, images, and more, through learning from existing data and pattern recognition. From another perspective, AIGC is a concept of general artificial intelligence, aiming to create AI capable of performing any human intelligence task. Within this framework, generative AI is one approach or pathway to achieve the goal of general AI, generating realistic data through adversarial training between a generator and a discriminator. Currently, common information carriers include text, images, audio, video, documents, web pages, and more \cite{wu2023multimodal}. AIGC refers to the use of AI technologies, particularly natural language processing and generative models, to automatically generate content/carriers in forms such as text, images, audio, and video. Some key technologies include natural language processing, generative models, text generation, image generation, audio generation, and video generation. The combination of these technologies enables AIGC to automatically generate various forms of content based on given conditions, expanding the application potential of AI in creative and media fields.

\textbf{1) Natural language processing (NLP)} \cite{Zhao2019asurvey}: It is a core technology for text processing, involving the conversion of human language into a form that computers can understand and process. This technology includes tasks such as text tokenization, part-of-speech tagging, syntactic parsing, named entity recognition, sentiment analysis, and more. By applying NLP to text, AIGC can understand and process the input language information to generate corresponding rich content.

\textbf{2) Generative models}: As an essential component of AIGC, generative models are a class of machine learning models that can generate text, images, audio, or video that adhere to specific rules based on given conditions. Representative algorithms for generative models include recurrent neural networks (RNNs) \cite{chauhan2018review}, variational autoencoders (VAEs) \cite{kingma2013auto}, and GANs \cite{goodfellow2014generative}. These models can learn the underlying distribution of data and generate content with logical and semantic coherence.

\textbf{3) Text generation}: Text is one of the most common information formats, making text generation one of the primary applications of AIGC \cite{li2024advances}. By training generative models, new text can be generated based on input text conditions. For example, given a textual description, GPT can generate articles, comments, news reports, and more in a similar style and content, which is known as "prompt engineering". Text generation has broad applications, including news summarization, automated writing, intelligent customer service, content creation, and more.

\textbf{4) Image generation}: It is another significant application of AIGC \cite{elasri2022image}, benefiting from the rapid development of computer vision technologies. Users can generate realistic images quickly based on given conditions, such as textual descriptions, semantic labels, style references, and more, using generative models. For example, one can input a textual description and generate an image that matches the description. Image generation techniques can be applied in e-commerce, advertising design, computer games, virtual reality, art creation, and other fields.

\textbf{5) Audio generation}: It is an important technology within AIGC \cite{liu2024cogeneration}. Through generative models, realistic audio can be generated based on given conditions, such as note sequences, audio samples, sheet music, and more. For example, a user can input a sequence of musical notes, and AIGC can generate corresponding audio melodies based on the given style. Currently, audio generation technology has been successfully applied in music composition, speech synthesis, and virtual assistants, but it has also been misused by malicious actors for illegal activities such as voice phishing.

\textbf{6) Video generation}: Video is a highly intuitive and easily understandable data format, but video generation is a complex technology. Through relevant AI technologies and generative models, realistic videos can be generated based on given conditions, such as scene descriptions, image sequences, action instructions, and more. For example, a continuous video can be generated from a set of image sequences or a realistic video can be generated based on a relevant instruction, such as the Sora software. Video generation technology can be applied in film special effects, virtual reality, video editing, and other fields.

\subsection{Applications and Potential of AIGC in LA}

The application of AIGC encompasses core components of the discipline of LA, and it holds great potential in this field. It can analyze vast amounts of data, generate designs, and simulate various scenarios, contributing to the design of sustainable, efficient, and high-quality landscape environments. The relevant applications cover various stages and key areas of landscape planning and design, plant selection and application, construction management, and process optimization. AIGC exhibits the following advantages in landscape architecture, including but not limited to:

\textbf{1) Accelerating the design process and improving efficiency through data processing and analysis}: AIGC can handle large-scale site data, climate data, plant databases, and more, providing comprehensive data support. It can automatically collect, integrate, and analyze data. It then can speed up the design process and offer more accurate data analysis results. AIGC reveals the relationships and trends among data, helping designers understand the impact of site conditions and environmental factors on the design. By predicting and simulating data, it assists designers in anticipating plant growth, lighting distribution, and other aspects, optimizing design solutions.

\textbf{2) Providing creativity and inspiration by generating innovative design solutions}: AIGC can generate entirely new design proposals with similar styles by learning from and imitating existing design works. It offers diverse design choices and innovative design elements, inspiring designers' creativity. Through automatically generated design sketches, conceptual models, and more, AIGC assists designers in quickly capturing inspiration and expressing design ideas. Therefore, AIGC expands design possibilities while stimulating designers' creativity and exploratory spirit.

\textbf{3) Supporting intelligent decision-making and user-friendly visual interaction}: AIGC can generate realistic landscape visualizations and animations, aiding designers and stakeholders in better understanding and evaluating design proposals. This helps improve the visualization of design information and enhances the attractiveness of proposals, leading to better communication. AIGC can create virtual reality (VR) \cite{burdea2003virtual} or augmented reality (AR) \cite{billinghurst2015survey} experiences, allowing designers and stakeholders to immerse themselves in design proposals. This immersive experience enhances understanding and engagement with proposals, facilitating better decision-making and feedback.

Here are the potential applications of AIGC in the field of LA, including but not limited to:

\textbf{1) Landscape design generation}: AIGC can generate landscape designs based on specific constraints, user preferences, and environmental factors. For example, machine learning algorithms can analyze past designs and generate optimized new designs. By inputting design maps, AIGC can generate landscape proposals using generative AI algorithms such as Stable Diffusion\footnote{\url{https://github.com/AUTOMATIC1111/stable-diffusion-webui}} and Midjourney\footnote{\url{https://www.midjourney.com}}.

\textbf{2) Landscape performance simulation}: AIGC can simulate the performance of different landscape designs, including their impact on water flow, soil, and microclimate. By using machine learning algorithms and historical data, it can predict the performance of different designs. In landscape planning projects, AIGC can simulate the impact of different green space designs on rainwater runoff. By analyzing factors such as vegetation types, soil permeability, and drainage systems, it can predict the efficiency of rainwater collection and the reduction of surface runoff for each design option. This helps compare and identify the best design solution to reduce urban flooding and water wastage.

\textbf{3) Environmental analysis}: AIGC can analyze the environmental impact of various landscape designs, such as carbon footprint, water resource usage, and biodiversity. By using machine learning algorithms, it can analyze environmental data and generate suggestions to improve design sustainability. For example, UrbanFlow developed by OYA Landscape Research is an intelligent design tool for urban renewal. Designers can select time frames and open space index (OSI) \cite{nega2010open} types to generate analysis maps of OSI density and mix. It helps designers understand the current distribution density, mix/diversity, and aggregation trends of different functional formats in urban spaces, providing valuable references for industrial planning and format layout.

\textbf{4) Public engagement}: AIGC can enhance public participation in the landscape design process by allowing them to interact with virtual models and provide feedback. By using appropriate machine learning algorithms, it can analyze public feedback and generate designs that better align community needs. For example, users or designers can incorporate public participation into the landscape design process through certain urban planning software tools to collect public opinions and resident needs, which is an important and indispensable part of landscape environmental assessment.

\textbf{5) Data visualization} \cite{chen2008brief}: AIGC can visualize complex landscape data, such as terrain, soil composition, and plants. By using machine learning algorithms, it can generate interactive and easy-to-understand 3D models \cite{remondino2006image}. For example, AIGC can be used to visualize complex urban elements such as buildings, roads, public facilities, and green spaces, helping users better understand the spatial distribution, population density, land use mix, and other characteristics of urban development proposals, and facilitate comparison and analysis of these features.

\textbf{6) Project management}: AIGC can manage and coordinate landscape projects by tracking progress, identifying potential issues, and generating schedules. By analyzing project data using machine learning algorithms, it can generate suggestions to improve efficiency and reduce costs. For example, the Shunyin Landscape Intelligence System \footnote{\url{https://www.shunyinzhichuang.com}} integrates landscape information modeling (LIM) to solve conflicts between specific construction and complex structures and professional integrated design, achieving high collaboration, accurate output of engineering quantity, and digital simulation of construction.

\textbf{7) Maintenance and monitoring}: AIGC can monitor and maintain landscapes by analyzing data from sensors and other sources. By using machine learning algorithms, it can predict maintenance needs, identify potential problems, and generate recommendations for improving the health and longevity of landscapes. Through relevant software tools, it monitors and maintains landscapes, analyzes data from sensors and other sources, predicts maintenance needs, identifies potential problems, and generates recommendations to improve the health and longevity of given landscapes.

\textbf{8) Urban planning} \cite{levy2009contemporary}: AIGC can be used for planning and designing urban landscapes by analyzing population density, transportation, and infrastructure data. By using machine learning algorithms, it can generate suggestions to improve urban sustainability and enhance the aesthetic environment. It can analyze data such as population density, land use mix, traffic flow, and socio-economic indicators, and generate recommendations to improve the livability, sustainability, and effectiveness of urban areas.

\textbf{9) Climate change adaptation}: AIGC can be used to plan and design landscapes that adapt to climate change. By analyzing climate data such as temperature, rainfall, and sea-level changes using machine learning algorithms, AIGC can generate recommendations for landscape sustainability. This includes mechanisms or technical approaches such as using local plants, green infrastructure, and water management systems to mitigate the effects of global and regional climate changes. Climate change is a severe and realistic challenge faced by all humanity.

\textbf{10) Interdisciplinary collaboration and innovation}: AIGC promotes collaboration and innovation between LA and other disciplines, such as geography, sociology, ecology, design, and AI engineering. It generates design solutions that consider multiple fields comprehensively. For example, intelligent layout generation is a key task of AIGC during the presentation of design proposals. It refers to the algorithms and models of AIGC that automatically generate or improve design layouts \cite{shi2023intelligent}. This field of research involves multiple disciplines such as computer vision, aesthetics, and AI, demonstrating significant interdisciplinary characteristics.

In summary, the potential applications of AIGC in LA include landscape design generation, landscape performance simulation, environmental analysis, public engagement, data visualization, project management, maintenance and monitoring, urban planning, climate change adaptation, and interdisciplinary collaboration and innovation. These applications leverage machine learning algorithms to analyze data, generate design proposals, simulate performance, engage the public, optimize project management, visualize data, monitor landscapes, plan urban spaces, adapt to climate change, and foster collaboration across disciplines.

\subsection{Key Technologies of AIGC in LA}

Natural language processing (NLP) and computer vision (CV) \cite{xu2021computer} are two fundamental technologies for deep learning in generative AI. In large AI models, which primarily operate in the NLP domain, the main techniques include speech processing, text mining, and text analysis. The Transformer architecture \cite{parmar2018image} has replaced recurrent neural networks (RNN) as the standard architecture for generative AI, and ChatGPT \footnote{\url{https://chatgpt.com/}} is a pre-trained model built on the Transformer architecture. In large visual models, which primarily operate in the CV domain, the aim is to enable computers to recognize objects, scenes, and motion in images, thereby enabling applications such as image search, image recognition, image restoration, intelligent monitoring, scene reconstruction, and augmented reality. The Vision Transformer (ViT) architecture \cite{khan2022transformers} has demonstrated more powerful capabilities compared to traditional CNN-based techniques. Building upon the general architecture, generative modeling techniques such as GANs and diffusion models can be used to achieve content generation and creation.

Note that the application of AIGC in LA is still in its early stages. There are already several designed or completed cases showcasing the auxiliary applications of AIGC technologies. AIGC is mainly applied in landscape planning and design, encompassing site surveys and analysis, design concept and scheme generation, parametric design and optimization, plant layout and simulation, construction management and optimization, and other aspects. For example, the technologies involved in the planning and design stages include parametric modeling and optimization, while the main technologies involved in the construction stage (e.g., construction drawing design) include BIM and LIM. In general, AIGC can provide technical applications in data preprocessing, image generation, text generation, management, and optimization in the field of landscape architecture, as shown in Table \ref{table: Key technologies}. The related platforms, systems, and software are listed in Table \ref{table: Related platforms}.

\begin{table*}[ht]
 \centering
    \caption{Key technologies of AIGC in LA}
    \label{table: Key technologies}
    \scalebox{0.75}{ 
    \renewcommand\arraystretch{1.2}
      \begin{tabular}{l|m{4cm}<{\raggedright}|m{17cm}}
        \hline
        \hline
        \multicolumn{1}{c|}{\textbf{Field}} & \multicolumn{1}{c|}{\textbf{Technology}} & \multicolumn{1}{c}{\textbf{Application in LA}} \\
        \hline
        \hline
        \multirow{3}{*}{\rotatebox{90}{Data preprocessing}} & Word embedding & It is used to map words in text to real vectors in a continuous vector space. Word embedding can capture the semantic information and contextual relationships of words. Therefore, word embedding technology has shown extensive application value in many application scenarios of natural language processing, such as automatic text classification, sentiment analysis, and machine translation. \\ 
        \cline{2-3} 
        & Image encoder & It is a model or algorithm used to convert image data into a vector representation. Its main function is to convert pixel information in complex images into fixed-length vectors for processing and analysis in ML or DL models. \\ 
        \cline{2-3} 
        & Data generation GANs \cite{figueira2022survey} & It is not limited to image generation, but can also effectively handle diverse data types, such as text, GPS positioning data, and mobile data. They can simulate and generate new data samples that are close to real data, so enriching the corpus. It can generate more data samples, which have a significant effect on improving the performance of downstream learning tasks. In addition, GANs also have powerful data conversion capabilities and can migrate the characteristics of one dataset to another dataset. It provides an effective way for data enhancement and domain adaptation. \\ 
        \hline
        \multirow{6}{*}{\rotatebox{90}{Image generation}} & Convolutional neural network, CNN \cite{padarian2019using} & It can automatically learn from massive images and extract abstract features, so it is widely used in landscape image information recognition, feature extraction and classification. \\ 
        \cline{2-3} 
        & Variational auto-encoder, VAE \cite{kingma2013auto} & VAE is an important type of generative model. It uses recognition models and autoencoders to generate images and uses the logarithmic maximum likelihood method to estimate parameters. \\ 
        \cline{2-3} 
        & Generative adversarial network, GAN \cite{goodfellow2014generative} & As an innovative method in the field of unsupervised learning, its architecture mainly consists of two networks: the generative network and the discriminative network. The two networks engage in a dynamic adversarial game during the training process, and continuously adjust their parameters iteratively to achieve a goal - that is, making it difficult for the discriminant network to distinguish whether the data output by the generating network is real data. GANs are particularly prominent in image generation, capable of producing highly realistic and indistinguishable synthetic images. \\ 
        \cline{2-3} 
        & DALL-E & It is the AIGC large model proposed by OpenAI for image generation tasks. Different from traditional image generation models, this model can generate images that match the description based on the text description input by the user. \\ 
        \cline{2-3} 
        & Google street view, GSV & Suitable for evaluating green plants' functional and visual effects at street level. The previously used remote sensing images cannot capture the urban horizontal plane and ignore the height of the plants themselves, which is more suitable for macro-level urban greening analysis \cite{ruangrit2004remote,faryadi2009interconnections}. \\ 
        \cline{2-3} 
        & Image style migration, ISM \cite{guo2023heritage} & In the image generation and rendering process of landscape design, in order to improve the transfer effect, image style transfer technology is often used, and a specific image style transfer model is developed or selected. \\ 
        \hline
        \multirow{5}{*}{\rotatebox{90}{Text generation}} & Transformer \cite{vaswani2017attention} & A model specifically designed to handle sequential data such as natural language. This model is suitable for tasks such as translation and text summarization. It is also unique in its ability to process the entire input sequence at once, without the need for batch processing. This feature has greatly promoted the development of advanced pre-training models such as BERT \cite{kenton2019bert} and GPT, allowing them to demonstrate excellent performance and efficiency when processing natural language data. \\ 
        \cline{2-3} 
        & Generative pre-trained transformer, GPT \cite{nyqvist2024can} & It uses an autoregressive generation method, which can generate subsequent text based on the previous text, and conducts pre-training through large-scale text data, thereby learning rich language knowledge and semantic information. It is used for text generation, dialogue systems, and language understanding. \\ 
        \cline{2-3} 
        & Bidirectional encoder representation from transformers, BERT \cite{kenton2019bert} & It is a pre-trained language model proposed by Google, which significantly highlights its innovation in language representation generation. It introduces a new masked language model (MLM), which enables the model to generate deeper and bidirectional language representations, thereby demonstrating excellent performance in language understanding and generation tasks. \\ 
        \cline{2-3} 
        & Recurrent neural network, RNN \cite{chauhan2018review} & It is a neural network architecture specifically designed to process sequence data. Its unique structure enables it to efficiently capture dynamic changes in data in a sequence. Therefore, RNN can accurately distinguish and generate corresponding output when processing sequence data with different contextual meanings, thereby effectively solving the problem of semantic differences caused by contextual differences. \\ 
        \cline{2-3} 
        & Long short-term memory, LSTM \cite{hochreiter1997long} & As a special variant of RNN, its main goal is to solve the challenges of gradient disappearance and gradient explosion encountered by traditional RNN when processing long sequence data. Compared with standard RNN, LSTM can show better performance on longer sequence data due to its unique structural design. \\ 
        \hline
        \multirow{2}{*}{\rotatebox{90}{Management}} & BIM / LIM \cite{nessel2013place} & It is a process, method, and technology that creates and uses digital models to manage and optimize the entire process of design, construction, and operation of landscape engineering projects. \\ 
        \cline{2-3} 
        & Digital twin, DT \cite{deng2021bim} & Prominent technology for instant two-way integration of cyber-physical systems and supporting intelligent decision-making. Emerging from BIM through four stages of development and integration with other technologies, involving different levels of the use of sensors, simulation, and AI \\ 
        \hline
        \multirow{4}{*}{{\rotatebox{90}{Optimization}}} & Genetic algorithms, GA \cite{guo2010enhanced} & It is a series of heuristic algorithms inspired by the theory of natural evolution. Similar to natural evolution, these algorithms provide excellent solutions to problems ranging from search to optimization to ML. Applying GA within Rhino Grasshopper software can effectively automatically identify the optimal layout of modules with different proportions and configurations, and then create the most appropriate environmental layout plan under given volume ratio conditions. \\ 
        \cline{2-3} 
        & eCognition \cite{zhang2018practice,zhang2020AnRS} & The emerging intelligent high-resolution remote sensing image analysis breaks through the limitations of traditional commercial remote sensing software that only performs image classification based on spectral information, and provides a new remote sensing analysis approach for landscape development. \\ 
        \cline{2-3} 
        & Machine learning \cite{alpaydin2020introduction,jordan2015machine} & It includes three aspects: data, algorithm, and application platform. It is good at summarizing patterns in a variety of data and solving different specific problems in the information extraction, analysis and evaluation, and planning and design stages of the landscape architecture workflow. \\ 
        \cline{2-3} 
        & Deep learning \cite{goodfellow2016deep,hinton2006fast} & It has powerful analytical capabilities for image feature extraction and has become a tool for automatically identifying land surface types \cite{wang2022Research}. It recognizes high-resolution remote sensing images, street view pictures, network media data, and PPGIS platform data used to assist in recognition. \\ 
        \hline
        \hline
    \end{tabular}}
\end{table*}

\begin{table*}[ht]
    \centering
    \caption{Related platforms, systems, and software}
    \label{table: Related platforms}
    \scalebox{0.75}{ 
    \renewcommand\arraystretch{1.2}
    \begin{tabular}{|m{2.5cm}<{\raggedright}|m{2.5cm}<{\raggedright}|m{13cm}|m{3cm}<{\raggedright}|}
        \hline
        \hline
        \multicolumn{1}{|c|}{\textbf{Platform}} & \multicolumn{1}{c|}{\textbf{Service Objects}} & \multicolumn{1}{c|}{\textbf{Introduction}} & \multicolumn{1}{c|}{\textbf{Website Link}} \\
        \hline
        \hline
        Spacemaker & Designers, real estate developers, government & An intuitive, collaborative, cloud-based AI software focused on generating high-quality site plans. It can complete feasibility analysis, generative design, visualization, solution comparison, and quantitative evaluation features. & \url{https://www.autodesk.com/products/forma} \\
        \hline
        Archistar & Designers, real estate developers, planning decision makers & Developed by an Australian company in 2018, It can complete land analysis, construction quality assessment, visual design, market forecast, etc. for real estate projects. Can quickly generate and modify design plans, achieve rapid design verification and optimization, analyze market trends, predict needs, and evaluate the competitive environment. & \url{https://www.archistar.ai} \\
        \hline
        XiaoKu Xkool & Public, designers, real estate developers & A web cloud design platform independently developed by Shenzhen Xiaoku technology. It integrates the construction engineering industry specifications, automatically generates intelligent designs, and establishes a building-type library and a unit-type library. It can complete intelligent design and evaluation of building types, information translation of land use conditions, indicator setting, automatic generation of layout plans, and intelligent rendering of general plans. & \url{https://cloud.xkool.ai}  \\
        \hline
        ShiFang DEEPUD & Public, designers, real estate developers, government & An urban design cloud intelligent design platform based on cloud computing and machine learning developed by Shanghai Chengsuan Information technology 2021. It uploads CAD, generates hundreds of design plans, inputs indicator parameters for design deduction, and can export master plans of different design styles, Renderings, analysis diagrams, plan VR tours, and Preview. & \url{https://www.shifang.city} \\
        \hline
        Scout & Public, designers & An interactive design system developed by KPF in the United States in 2016 based on the parametric software of Rhino and Grasshopper. It can be operated on the web page. It is a generative design platform based on a parametric technology framework and supported by the computing power of a single workstation. & \url{https://ui.kpf.com}  \\
        \hline
        Delve & Designers, real estate developers, government & A web-based operating platform developed by Google Alphabet in 2019, and its working principle is similar to Scout. It can comprehensively sort the various generated plans according to the evaluation level, generate a more detailed estimate of construction income, and produce a plan design that meets specifications and is economically feasible. & \url{https://www.sidewalklabs.com/products/delve} \\
        \hline
        UrbanFlow & Designers & An intelligent design tool for urban renewal neighborhoods launched by Oya in 2024. It integrates multiple technologies such as parametric, AI, and geographical information systems, and has five core modules: urban base generation, urban data analysis, road structure adjustment, node space simulation, and scene atmosphere generation. & \url{https://urbanflow.co} \\
        \hline
        ShunYin ZhiKu & Designers & An intelligent landscape system that can realize intelligent landscape design, intelligent landscape planning, and LIM. With the advantages of Revit software, it can achieve high professional collaboration, accurate and direct project quantities, digital simulation of construction, etc. & \url{https://www.shunyinzhichuang.com} \\
        \hline
        Midjourney & Designers & As an AI drawing platform integrated into Discord, it works by generating images from natural language descriptions. You can choose different artistic styles and input prompt words to generate corresponding images through AI algorithms. & \url{https://www.midjourney.com} \\
        \hline
        Stable Diffusion & Designers & A deep learning model for converting text into images. It generates high-quality, photo-realistic images from any text input. It uses the CLIP ViT-L/14 text encoder, which can adjust the model through text prompt words. The model itself and the code used to generate the images are completely open source. & \url{https://github.com/AUTOMATIC1111/stable-diffusion-webui} \\
        \hline
        SUAPP & Designers & A SketchUp extension plug-in. It is not a traditional renderer that renders through physical simulation or ray tracing. Instead, it uses AI technology to process various forms of input data. Users can quickly generate high-quality design previews by simply adjusting parameters. Pictures, including models, line drawings, on-site photos, etc. & \url{https://check.suapp.com/plugin/539} \\
        \hline
        Grasshopper & Designers & A plug-in that runs in Rhino, it can build model logic by connecting operators to generate models based on program algorithms. It uses the programming capabilities of GPT to query, analyze, and call the programming codes of Rhino during the design process, greatly improving the efficiency of modeling and design when the special-shaped landscape design is deepened. & \url{https://www.grasshopper3d.com/} \\
        \hline
        Digital Blue Foam & Designers, real estate developers, government & An AI platform developed by a Singaporean company, with advanced AI algorithms and data integration, it can realize functions such as derivative design, spatial analysis, intelligent optimization, and uses multi-source data to run analysis and realize real-time BIM functions. & \url{https://www.digitalbluefoam.com} \\
        \hline
        Sefaira & Designers & A tool developed by Trimble for early analysis. It integrates with SketchUp to perform energy performance and solar analysis, model simple geometries, analyze and compare massing, layout, and envelope options, and shortlist preferred concepts. & \url{https://www.sketchup.com/products/sefaira} \\
        \hline
        VIM & Designers & It converts complex BIM data into actionable real-time analysis. Through VIM cloud which provides dynamic real-time power BI data reporting, and software applications such as VIM enterprise that support BIM data management and customization, it provides real-time interactive 3D visualization for Construction professionals and offers new ways to manage and utilize data. & \url{https://www.vimaec.com} \\
        \hline
        Environmental Insights Explorer & Designers, government & It uses Google map data and advanced analysis technology to predict urban environmental data, mainly covering carbon emissions, air quality, etc. Among them, Tree Canopy combines AI technology with urban aerial images. It can help decision-makers better understand urban environmental data and formulate more effective strategies. & \url{https://insights.sustainability.google} \\
        \hline
        DreamzAR & Public & An online AI-based application designed to generate design plans and create 2D and 3D landscape designs. It can analyze visual information in photos to identify courtyards, add new landscape elements to them, and generate visualizations of exclusive garden and interior design plans. & \url{https://www.dreamzar.app} \\
        \hline
        \hline
    \end{tabular}}
\end{table*}

\section{Applications of AIGC in LA} \label{sec:applications}

\begin{figure*}[ht]
    \centering
    \includegraphics[width=17.2cm]{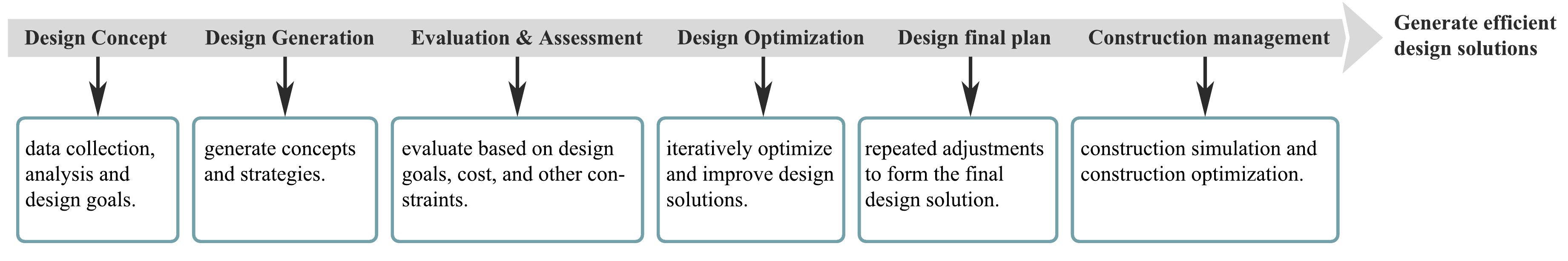}
    \caption{LA design process based on AIGC.}
    \label{fig:process}
\end{figure*}

LA design based on AIGC is embodied in the whole process of LA design. The design concept carries out data retrieval and collection, adopts algorithms to classify and discover data, and formulates design goals; generates design concepts and strategies; builds algorithm models; evaluates the design plan, cost, and other constraints; designs optimization, iterative improvement of the design plan; repeated debugging to form the final design plan; conducts construction management through construction simulation and construction optimization, as shown in Figure \ref{fig:process}.

\subsection{Site Research and Analysis}
\begin{figure}[ht]
    \centering
    \includegraphics[width=8.2cm]{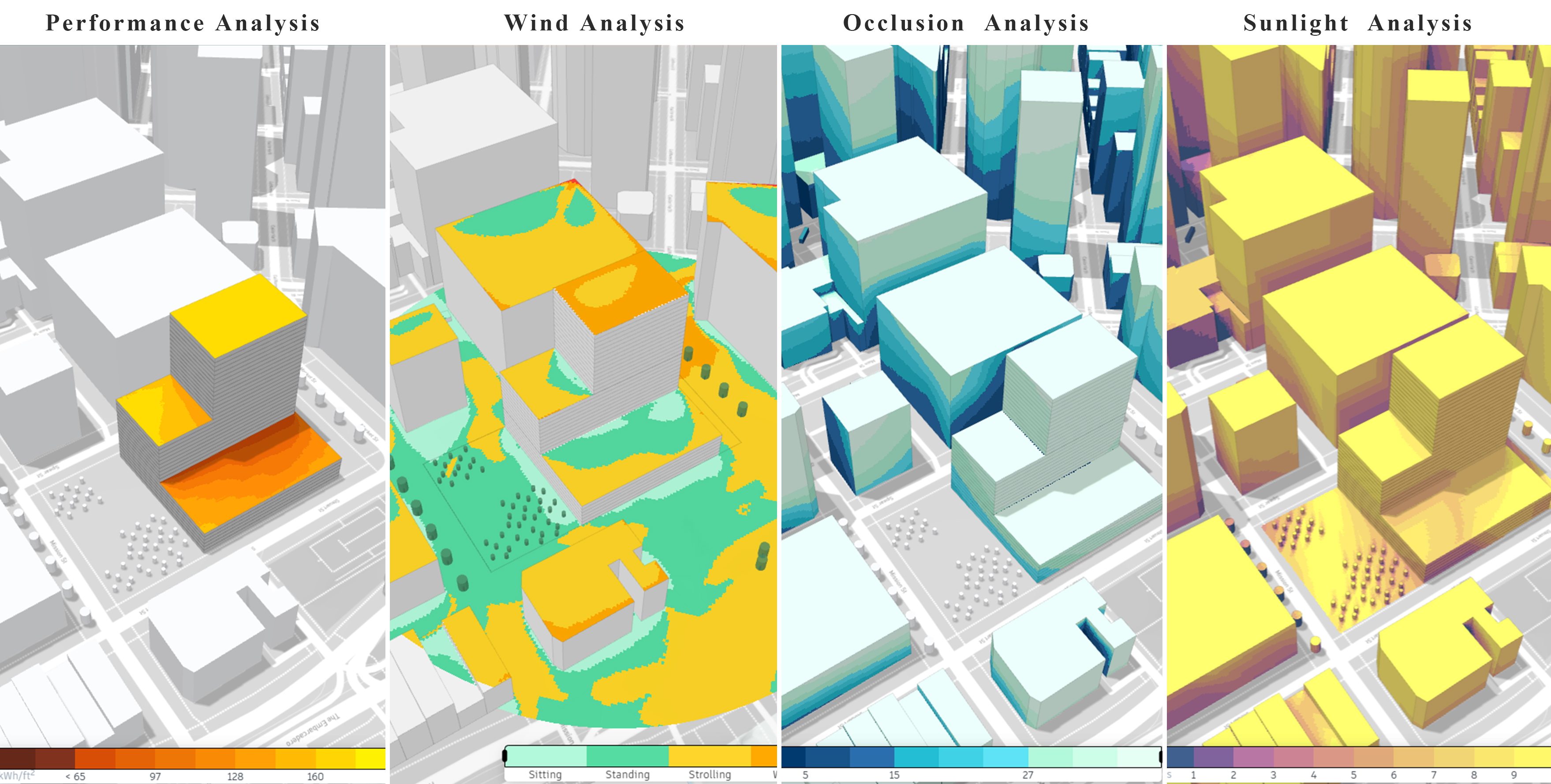}
    \caption{Site analysis}
    \label{fig:Site analysis}
\end{figure}

\textbf{1) Site analysis}. \textbf{a) Site analysis and evaluation}: AIGC can generate terrain models and visualization results by processing and simulating terrain data, enabling a better understanding of terrain characteristics for site evaluation. Traditional landscape analysis methods often struggle to accurately capture the continuity and heterogeneity of terrain data. By employing conditional GANs \cite{mirza2014conditional}, AIGC can analyze and process terrain data, acquiring deep spatial knowledge of the terrain through training \cite{zhu2020spatial}. This knowledge can be applied to terrain interpolation and analysis, resulting in more prominent interpolation results that demonstrate the effectiveness and superiority of AIGC in site analysis tasks. A tree study in Taipei \cite{wei2022artificial} utilized the i-Tree Eco model \cite{nowak2018tree} and self-organizing map algorithms to identify and assess urban trees, including the correlation between tree indicators and ecosystem services. It investigated the impact of different indicators on ecosystem services. The study found that AIGC can effectively identify relevant indicators of tree vegetation, leading to improved assessment and management of the urban landscape ecosystem. Face and object detection model techniques can analyze and interpret relevant images capturing human flow, dynamics, and behavioral patterns on social media platforms, which can be used to evaluate the usage of park green spaces \cite{song2022analyze}. \textbf{b) Solar radiation analysis}: AIGC can simulate solar radiation paths and lighting variations to analyze solar radiation data, predicting the intensity and distribution of sunlight during different periods and seasons. Conducting solar radiation analysis in the early stages of design is crucial for landscape form design and optimization. Using CNN \cite{gu2018recent,isola2017image} and GANs \cite{goodfellow2014generative} can predicte solar performance based on plan drawings, constructing non-linear models to forecast overall solar performance indicators. These indicators serve as targets for performance-oriented landscape design optimization \cite{he2021predictive}. The study demonstrates the advantages of AIGC in image processing, saving the complexity of parameter input and computational efficiency. Autodesk's Forma \footnote{\url{https://www.autodesk.com.cn/support/technical/product/forma}} runs on a web-based platform, which can simulate various environmental data of the project site through AIGC and generate visualized site analysis. It can analyze sun exposure and help designers gain a deeper understanding of site information and make more comprehensive and sustainable design choices, as shown in Figure \ref{fig:Site analysis}.

\textbf{2) Data integration and analysis}. AIGC can integrate and analyze large amounts of site data, pedestrian flow data, climate data, plant data, etc., providing comprehensive data support and analysis results. Through data mining techniques \cite{gan2021explainable,gan2017data,gan2021utility} and machine learning techniques \cite{jordan2015machine}, AIGC reveals correlations between data, helping designers make data-driven decisions. AIGC-enhanced datasets encompass meteorological data such as temperature, humidity, and rainfall, as well as multi-dimensional environmental factors including satellite imagery, land cover classification, and soil texture. This provides strong technical support for environmental data analysis by designers. It can enrich the dimensions of data and significantly improve the accuracy and efficiency of data analysis. For data extraction, Google Street View (GSV)\footnote{\url{https://www.google.com/streetview/}} and Tencent Street View (TSV)\footnote{\url{https://lbs.qq.com/tool/streetview/}} image extraction, as well as geotagged photos from online photo-sharing services like Flickr and panoramic photos from Panoramio, have become major data sources for urban research \cite{rundle2011using}. In a study of tourist attraction areas in Seoul, street image information was collected through GSV, and a CNN-based algorithm was retrained to classify tourism photos. The final model was applied to the entire dataset \cite{kang2021transfer}, demonstrating the potential of GSV as an effective data source for urban streets \cite{rundle2011using}. Compared to remote sensing images capturing macro-level urban greening analysis, evaluation, and visualization from sensors \cite{faryadi2009interconnections,ruangrit2004remote}, GSV captures the street level and the height of plants themselves, making it suitable for evaluating the functionality and visual effects of green plants on street level \cite{anderson2018estimating}. However, the extraction of street view greening images may inevitably have errors, such as when RGB color recognition confuses plant shadows with artificial colors in the street (e.g., billboards and wall surfaces). Generative adversarial network (GAN)-based models optimize satellite image resolution and address cloud cover issues. Satellite imagery has shown its indispensable value in large-scale landscape planning \cite{abdi2020land}. However, open data resources often suffer from low image resolution and cloud cover issues, requiring manual visual analysis to obtain data information \cite{krizhevsky2017imagenet}. Using GAN technology, clouds can be removed from satellite images, enabling high-accuracy building detection \cite{ikeno2021enhanced}. GANs can also optimize and enhance the spatial resolution of satellite images. Landsat satellite images, which are freely available, can be trained using four common reflectance bands (red, green, blue, near-infrared) and the panchromatic band. Through optimization using perceptual loss functions, higher spatial resolution satellite images are generated. This enables time-series classification and monitoring of land cover \cite{pham2021spatial}.

\textbf{3) Decision support}. Decision support in LA includes design option selection, resource allocation, and time planning, among others. AIGC can provide decision support and risk management strategies for designers, reducing the risks and uncertainties associated with the design process. It can simulate human visual perception and landscape features to help designers assess trends in land changes, predict land use, and evaluate the visual effects of different landscape design options. Decision support can offer multiple evaluation indicators and methods, such as multi-criteria decision-making methods and model simulations. By generating virtual scenes and visual simulations, the impact of different elements and materials on landscape aesthetics can be predicted, providing references for visual decision-making. Satellite imagery visualization can be used to understand and assess trends and impacts of land changes. By retrieving layouts from historical satellite images of cities and embedding them into neural networks, we can infer urban development patterns and predict spatial patterns of future urban growth. A study \cite{ibrahim2023expatriates} used GANs to predict future urban growth maps for Doha, visually illustrating the distribution of different population groups. By considering natural evolution trends in land use, socio-economic factors, policy indicators, and other variables, the research enables an intuitive understanding and assessment of land use changes, their impacts, and the relationship between humans and the environment. Establishing simulation models for land use and land cover changes is of great significance for understanding the process of urban change and evaluating the ecological landscape of the land \cite{verburg2004land}, providing planning and management strategies for LA design. Regarding the prediction of land use and land cover changes (LUCC) using pix2pix (GAN) \cite{isola2017image}, a study trained a nested U-Net model by learning the land use classification results and planning information of subway stations in Shenzhen from 1988 to 2015. It expanded the binary land use results into probability maps and used Landsat imagery for semantic segmentation, resulting in land use results with an accuracy exceeding 91\% \cite{sun2021gan}. Experimental results demonstrate that AIGC's technology can accurately predict future land use changes, providing important decision support for land landscape planning and management.

\textbf{4) Environmental optimization and risk assessment}. \textbf{a) Environmental optimization}: By integrating ecological principles into AI design, environmental optimization can help preserve existing vegetation, create wildlife habitats, and manage water resources, thereby improving the ecological quality of the landscape. Based on environmental data and simulation results, AI can provide recommendations and solutions for environmental optimization. For example, based on climate conditions and plant requirements, AI can recommend the most suitable plant species and layout to improve plant growth and landscape effects. Additionally, AI can optimize water resource utilization, energy consumption, and other aspects to help designers achieve environmental sustainability and energy-saving goals. If a designer needs to optimize the environment of a park, they can choose suitable plant species, increase green coverage, and improve air quality. Rainwater collection systems can be designed to use collected rainwater for irrigation, achieving sustainable utilization of water resources. Wetlands and water features can be designed to serve as habitats and water sources, fostering ecosystem restoration and protection. \textbf{b) Risk assessment}: By analyzing environmental data and conducting simulations, AI can help designers evaluate potential environmental risks and hazards. Risk assessment applications in LA design mainly include natural disaster risk assessment, ecological environment risk assessment (including vegetation destruction, soil erosion, water pollution, etc.), and facility equipment risk assessment (such as amusement facilities, bridges, lighting systems, etc.). By assessing the impact of design options on the ecological environment, AI can identify potential environmental risks and propose corresponding environmental protection strategies, such as vegetation restoration and ecological wetland construction. In urban park planning and construction, it is necessary to assess the natural disaster risks faced by the park, such as floods and earthquakes. Through risk assessment, the vulnerability of design options to natural disasters can be evaluated, and appropriate measures can be taken to mitigate risks, such as selecting seismic-resistant materials and optimizing drainage systems. By using GIS and historical data, potential risk areas can be identified, and park layouts and architectural designs can be adjusted accordingly to reduce disaster risks. AI can predict the likelihood and impact of natural disasters such as earthquakes, floods, droughts, and storms, providing designers with corresponding response measures and prevention strategies to ensure the safety and sustainability of LA design. Convolutional neural networks (CNN) \cite{krizhevsky2017imagenet} and semantic segmentation are effective tools for processing street view images to determine building heights, enabling the assessment of earthquake risks in cities. New machine learning models, such as visual transformers or their semantic segmentation versions, have made significant progress in image classification tasks \cite{urena2023automatic}. A study \cite{kuccukdemirci2023investigating} in ancient agricultural fields used a convolutional neural network to detect and automate analysis of the archaeological features hidden in the forests of the Swedish landscape. For this purpose, the study used a U-shaped CNN architecture called U-net \cite{ronneberger2015u} to detect/extract clearance cairns from the LIDAR dataset. The results demonstrated the model’s ability to identify ground anomalies, aiding in the recognition of large-scale complex ground environments and contributing to disaster warnings and risk assessment.

\subsection{Design Concept and Plan Generation}

\begin{figure*}[ht]
    \centering
    \includegraphics[width=0.9\textwidth,height=0.6\textwidth,scale=1.0]{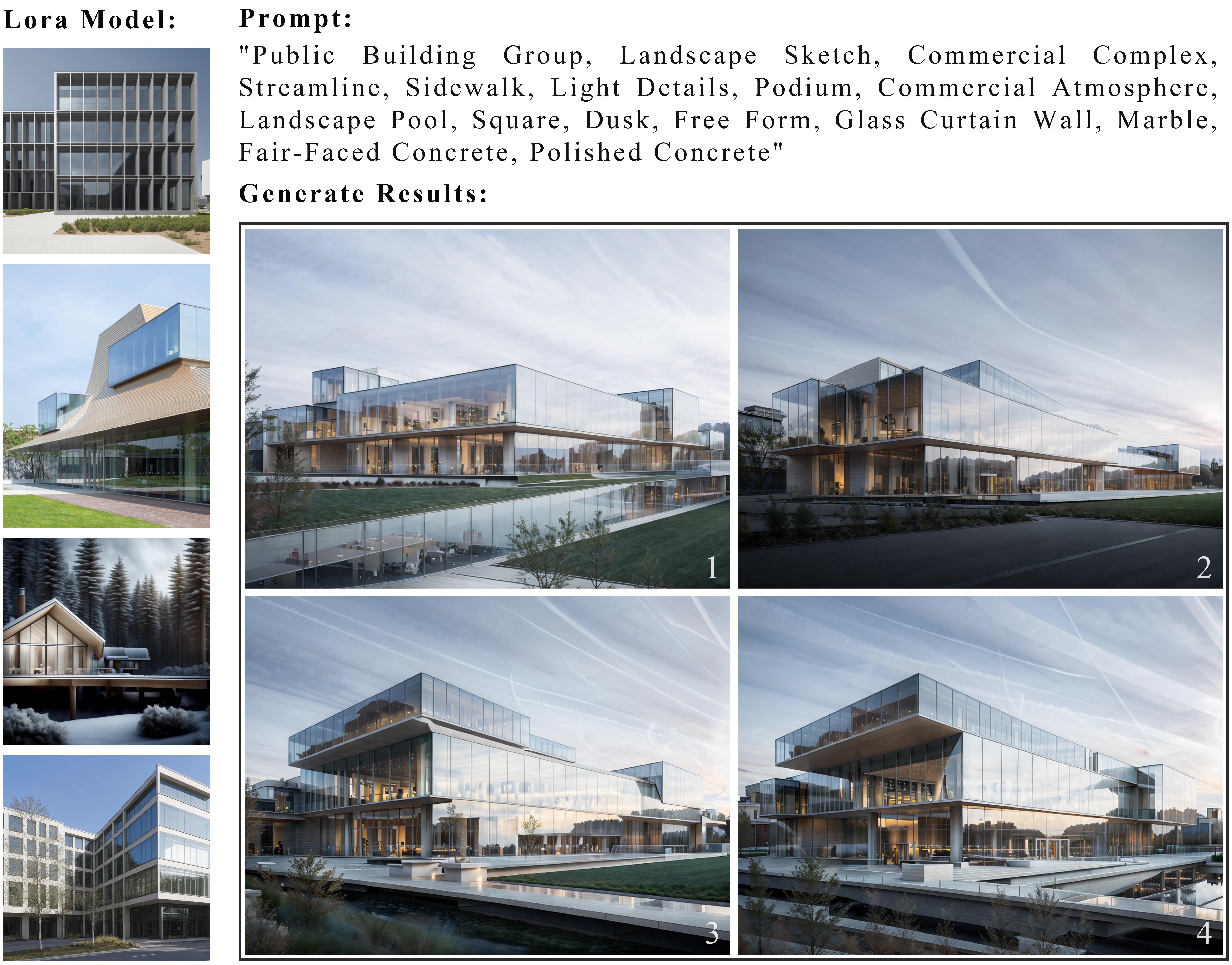}
    \caption{Image synthesis}
    \label{fig:image synthesis}
\end{figure*}

\textbf{1) Visual analysis and image synthesis}. AIGC plays an important role in visual analysis and image synthesis. Visual analysis refers to the process of analyzing and understanding visual data such as images and videos using AI techniques, while image synthesis is the process of generating new images or modifying existing ones using AI. Through deep learning algorithms and big data analysis, AIGC interprets on-site image data, identifies key elements such as urban textures, street scene elements, vegetation types, terrain features, and water layouts, and extracts their three-dimensional shapes, textures, and lighting information to achieve efficient analysis and precise reconstruction of landscape elements. Building upon visual analysis, AIGC technology can automatically generate three-dimensional models, providing visual experiences for designers. In terms of visual analysis, AIGC can be applied to tasks such as image classification, object detection, semantic segmentation, and instance segmentation. Deep learning techniques \cite{hinton2006fast} have become increasingly powerful tools for automatic recognition of land surface types due to their strong analytical capabilities in extracting image features \cite{wang2022Research}. High-resolution remote sensing images and street view pictures recognized by deep learning can be used for informal green space layout recognition and assisted recognition through network media data and PPGIS platform data \cite{brown2014key}. By constructing deep learning models, images can be semantically segmented, meaning that semantic units are used to segment the image into regions with different semantic labels, ultimately obtaining results of image region classification \cite{durduran2015automatic}. Compared to manual visual interpretation, deep learning models can analyze large amounts of data and learn from them, resulting in relatively objective recognition with higher accuracy, precision, and efficiency \cite{Zhang2023research}. In terms of image synthesis, AIGC can be applied to tasks such as image generation, style transfer, and image restoration. Through techniques like GAN \cite{goodfellow2014generative}, AIGC can generate realistic images and even images that are completely different from the real world. The application of AIGC in visual analysis and image synthesis has broad prospects and can provide more intelligent and efficient solutions in various fields. AIGC can be applied to synthesizing street view images. In the field of urban landscapes, complex analysis and data-driven simulations are required. In related cases, deep learning and image transformation techniques have been applied to landscape simulation in urban redevelopment projects. Kikuchi et al. \cite{kikuchi2023development} proposed a method that integrates Image-to-Image (I2I) technology \cite{zhu2017unpaired} and procedural modeling to automatically generate synthetic datasets for training instance segmentation models. By using real city 3D models and image transformation techniques, synthetic street view images capturing real-world landscape features can be generated. This can visually demonstrate the landscape changes before and after the project and provide strong support for landscape analysis. AIGC can also synthesize images with creative and visual characteristics, often using large datasets of street view images and satellite images for efficient and accurate analysis and application. A study \cite{steinfeld2022imaging} trained StyleGAN and Pix2Pix. StyleGAN generates images with dominant visual characteristics of urban environments by extracting feature combinations from latent space. Pix2Pix transforms given source images (depth maps of urban landscape spaces) to generate new synthesized images. On stable diffusion, the process involves several steps. Firstly, multiple landscape architecture images that share similar elements such as architectural style, form material, and environmental atmosphere are uploaded as the training dataset for the model. Secondly, based on the description of the required training set type, training direction, and other basic parameters, a precise and dedicated Lora model can be trained. Finally, through this model, parameters such as the width, height, batch size, and quantity of the generated images can be set, and accurate and realistic training results can be generated based on the required descriptive prompts. Additionally, to obtain better image quality, multiple trained Lora models can be stacked and used, with the weights of each Lora model adjusted to achieve optimized generation effects focusing on different aspects of the image. By inputting prompts optimized by ChatGPT, diverse and high-quality combinations of image parameters can be generated, resulting in images that are highly similar to real landscape environments and rich in creativity, as shown in Figure \ref{fig:image synthesis}.

\textbf{2) Landscape scheme generation}. AIGC can generate new design schemes by learning from and imitating existing LA designs. By analyzing and learning design elements, layouts, and organization, AIGC can generate innovative landscape design schemes that provide inspiration and references for designers. This process involves the integration of various technologies, such as machine learning combined with other techniques, deep learning, CNN, generative models, image recognition, image transfer, and more. Examples of these techniques include conditional GANs \cite{mirza2014conditional}, flow-based models \cite{kang2008flow}, diffusion models \cite{ho2020denoising}, Transformer \cite{vaswani2017attention}, variational auto-encoders (VAE) \cite{kingma2013auto}, and others. As a graph neural network (GNN) framework, Graph2Plan \cite{hu2020graph2plan} aims to automatically generate floor plans under user guidance. With this framework, users can specify a series of constraints such as positions, adjacency relationships, and building boundaries, and Graph2Plan can generate highly customized and accurate floor plans that satisfy these conditions. The convolutional message passing neural network (Conv-MPN) \cite{zhang2020conv} is used to efficiently perform higher-order inference, combining layouts and validating adjacency constraints \cite{nauata2020house}. Furthermore, image style migration (ISM)  can be used for image generation and rendering. ISM is a semantic segmentation-based algorithm that extracts semantic information from content and style images using masked R-CNN. To improve migration effectiveness and quality, an image style migration model for landscape design can be established by incorporating multi-scale discriminators, content feature mapping modules, and transfer learning methods \cite{guo2023heritage}. Research in this area focuses on image style migration modeling and its integration into urban landscape design. Future research areas in ISM include visual quality assessment, qualitative and quantitative evaluation of results, model scalability and generalization, and the integration of style migration into landscape design \cite{Gong2024Chinese}. Semantic-guided landscape image generation based on generative adversarial network models has been explored in a study that employed Swin Transformer text encoders to capture landscape art requirements in rural construction. The corresponding landscape images for rural construction were generated using GAN \cite{wang2023environmental}. Another approach is the local category-specific and global image-level GAN \cite{tang2020local}, which is guided by semantic maps and focuses on generating objects of different categories by building and learning multiple sub-generators. This approach effectively captures and represents scene details. Conditional GAN \cite{mirza2014conditional} also can generate synthetic images with high quality and semantic consistency, enabling independent control of style for each semantic region. Therefore, AIGC has significant advantages in generating landscape schemes based on semantic guidance. It can generate images of indoor scenes, rural landscapes, and urban street views, as well as perform cross-view image translation, producing high-quality and realistic synthesized landscape images. GAN-based models can also create textures and images in various styles for urban landscapes and building facades. QuiltGAN \cite{arantes2019quiltgan} is a texture generation algorithm based on conditional deep convolutional GAN \cite{luo2021case}. It generates high-resolution images by training a generator and a discriminator. The generator generates image blocks based on conditional image regions, and the discriminator evaluates the authenticity of the generated samples with respect to the conditional image. This ensures seamless integration of the generated image blocks with the surrounding environment while maintaining randomness in the generation process. QuiltGAN can generate nearly infinite variations from a single input image and performs well on various image types, including urban landscapes, building facades, and textures while preserving important large-scale constraints. CycleGAN \cite{sun2022automatic} can analyze and learn the style of historical architecture and generate stylized facades in traditional architectural styles. Using semantic segmentation techniques, it visually analyzes different styles of building facades in the Harbin Central Street architectural dataset. Then, through deep learning models, it learns and transforms the mapping relationships between different styles. The model can identify and extract key features of facade elements, thereby generating new architectural facades with specific styles. This technology not only improves the efficiency and accuracy of image generation but can also be used for the preservation of historical architectural heritage and urban landscape restoration, providing decision support for urban renovation and landscape planning. Based on image recognition methods, a study \cite{zhang2021intelligent} conducted intelligent image recognition of decorative windows in traditional private gardens in Suzhou, generating concepts in the urban scope using machine learning.

\textbf{3) Landscape scheme evaluation and assessment}. In landscape scheme evaluation, AIGC can utilize methods such as big data analysis, simulation techniques, and machine learning. It incorporates deep learning \cite{goodfellow2016deep,lecun2015deep}, such as deep neural networks (DNN) \cite{sze2017efficient}, Grasshopper platform \cite{sadeghipour2013ladybug}, generative adversarial networks (GAN) \cite{goodfellow2014generative}, and other models/algorithms, to enable rapid evaluation and optimization of design schemes, providing scientific and efficient support for landscape planning and design. AIGC can predict and evaluate the effects of different design schemes in real environments. By analyzing user feedback and behavioral data, it can assess the attractiveness and acceptability of design schemes. It can also use environmental simulation techniques to evaluate the impact of landscape design schemes on the surrounding environment. AIGC can comprehensively evaluate landscape design schemes considering various factors, including aesthetic effects, functionality, sustainability, and economic aspects. By considering factors such as energy utilization, carbon emissions, and water resource management, AIGC can assist designers in assessing and improving the sustainability of design schemes. It accomplishes this by establishing a multidimensional evaluation index system, adjusting and optimizing parameters, conducting comprehensive comparisons and evaluations of different design schemes, and identifying the optimal design scheme. AIGC can perform sustainability assessments and establish environmental assessment models to analyze the sustainability and environmental impact of design schemes. To address the evaluation of urban morphology, an automated urban morphology generation system called DeepCity \cite{Liu2023Artificial} has been developed by combining Python programs with the Grasshopper platform. DeepCity enables the entire process of case selection, data annotation, model training and generation, design evaluation, and data vectorization to be automated. Users only need to adjust some initial parameters in Grasshopper, reducing the threshold and time cost for designers. DeepCity possesses the capabilities of design cognition, design generation, and design evaluation. It can be used to evaluate and quantify different urban morphologies, conduct typological research on urban morphology, automatically learn morphological patterns of urban textures, apply them to new urban environments, and weave and repair specific types of urban morphology. It can also quickly assess the physical performance of design schemes, assisting in modifications and deepening of the schemes. Through GANs \cite{goodfellow2014generative}, it is possible to compare landscape plans created by landscape designers with landscape plans generated by generative design models. This allows for the evaluation of the authenticity and performance of the generated design models. Optimization objectives such as volume, layout, and architectural structure can be visualized, and filtered and optimized options can be generated. This provides technology-driven analysis and evaluation in the field of landscape design \cite{park2024quality}. An environmental landscape design and planning evaluation system based on deep learning (DL) and deep neural networks (DNN) has been developed \cite{chen2023environmental}. This system utilizes computer vision and deep learning techniques to analyze and evaluate the effects of landscape design. It automatically identifies and analyzes various elements and features of landscape schemes, including the environment, landscape level index, landscape distribution, landscape rationality, land use change management, and environmental impact. The evaluation results can be fed back to the previous design process for modifications. This helps designers better understand and meet user requirements, providing landscape design schemes that are functional, aesthetically pleasing, and environmentally sustainable. In terms of landscape design scheme evaluation, machine learning combined with other technologies has been applied to evaluate public green space satisfaction through social media. Text and image data are analyzed to understand public opinions, sentiments, or social phenomena related to urban fairness and satisfaction with public spaces (green areas). For example, social media and machine learning techniques have been used for evaluating satisfaction with parks in Beijing \cite{wang2021fine}, facilitating sustainable decision-making for urban green spaces. AI technology, with its efficient data processing capabilities and implicit rule-capturing abilities, is gradually being applied to landscape analysis and evaluation, replacing some simple repetitive tasks. Deep learning-based generative design further harnesses the creative potential of AI.

\subsection{Parameterized Design Optimization}

Parameterized design optimization in AIGC refers to the use of AI techniques in generative computing to automatically generate and optimize design solutions through parameterized modeling and optimization algorithms. This approach takes into account various parameterization factors in the design process and adjusts and optimizes these parameters using intelligent algorithms to generate optimal design solutions. Parameterized design is a new design form that is based on computer-aided design technology and combines variables and geometric constraints. The key elements of parameterized design are "parameters" and "algorithms", where parameters are the variables that should be defined in the design process, and algorithms are fundamental to the parameterized design. Designers set up the algorithm and logic of parameterized design based on design goals and defined variables, and ultimately modify the design variables to obtain corresponding design results. Machine learning is one of the core technologies of AI, enabling computers to automatically learn and improve from data. In landscape design, machine learning can be used to model and optimize various design parameters, such as layout and material selection, to achieve optimal design solutions. Parameterized modeling, represented by Rhino and its plugins \cite{de2020modeling}, greatly enhances the precision of work, enables visual programming modeling, and allows for modification of model results by modifying preset parameters.

\begin{figure*}[ht]
    \centering
    \includegraphics[width=\linewidth,height=0.7\textwidth,scale=1.0]{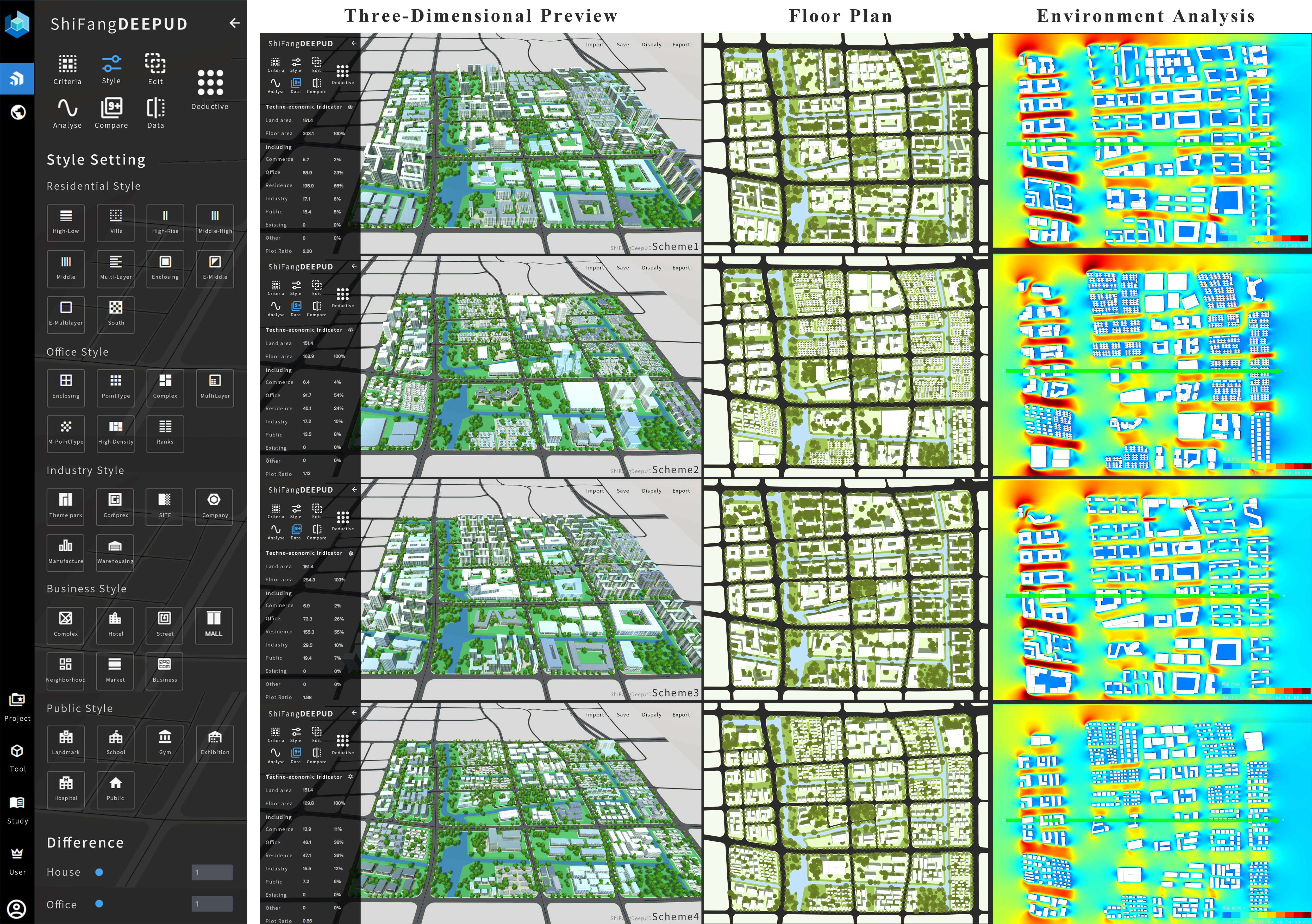}
    \caption{Parametric modeling}
    \label{fig:Parametric modeling}
\end{figure*}

\textbf{1) Parameterized design modeling}. This involves representing various factors and elements in the design process, such as terrain, vegetation, water bodies, and artificial structures, as adjustable parameters. By abstracting these elements into adjustable parameters, mathematical models of design solutions can be established. Parametric modeling encompasses the following application phases: \textbf{In image generation}, AI image generators such as DALL-E2\footnote{\url{https://openai.com/index/dall-e-2/}}, a text-to-image system introduced by OpenAI, or Stable Diffusion\footnote{\url{https://github.com/AUTOMATIC1111/stable-diffusion-webui}}, an advanced text-to-image model based on the Amazon Bedrock platform, can be used. These generators can create complex artworks using basic natural language prompts and generate conceptual renderings from user-generated text strings. \textbf{In design solution generation}, landscape designs can be generated based on design-based maps using generative AI algorithms. Stable Diffusion, Midjourney, and other generative AI tools operate on this principle. The core of Stable Diffusion is diffusion models, which use the CLIP ViT-L/14 \cite{radford2021learning} text encoder to adjust the model based on textual prompts. It separates the imaging process into a "diffusion" process starting from a noisy state, gradually improving the image until there is no noise, and progressively approaching the provided textual description. \textbf{In automatic solution evaluation}, two neural network methods are employed to assess whether a solution meets the required standards. In the two-stage evaluation model, the neural network first segments the semantic content of the solution to construct a graphical model and then completes the evaluation through graph classification. In the one-stage evaluation model, an end-to-end graphical model is built with a neural network, enabling graph classification and solution evaluation. \textbf{In automatic recommendation and adjustment of solutions}, generated solutions may contain design errors that do not adhere to design criteria. For solutions with low evaluations, a recommendation model is utilized to adjust the solutions based on modified topologies, resulting in more reasonable design solutions. \textbf{In intelligent solution generation}, the algorithm takes into account the suitability of plot sizes and can adjust and deform individual elements to match the target space. The Rhino Grasshopper plugin\footnote{\url{https://www.grasshopper3d.com/}} leverages the programming capabilities of GPT to query, analyze, and invoke programming code in Rhino Common and Grasshopper during the design process, providing rapid support for complex landscape designs. ShiFang DEEPUD\footnote{\url{https://www.shifang.city/}} utilizes machine learning to conduct image learning on urban form proposals, establishing an urban spatial database to provide more accurate design proposals. The process begins with importing the project's geographical data and setting the basic parameters of the site, such as plot boundaries, land use types, plot ratio, and building density. Next, based on the project requirements, different design styles corresponding to different land use types are customized. Then, using the set styles and indicators, intelligent deduction of design proposals can then begin, generating multiple potential design options based on the built-in model database. Finally, designers can preview the effects of multiple design proposals on the interface and select the most suitable proposal for further editing. The software provides quantitative data on different proposal indicators, master plans, and visualized environmental analysis results, enabling designers to choose more accurate, innovative, and sustainable solutions, significantly improving the efficiency and quality of urban design. Details are shown in Figure \ref{fig:Parametric modeling}.

\textbf{2) Design optimization algorithms}. Utilize optimization algorithms such as genetic algorithms, particle swarm optimization, simulated annealing, etc., to adjust and optimize the parameters in the design model. Through continuous iteration and optimization, the optimal design solution is sought. An experiment using an improved genetic algorithm \cite{guo2010enhanced} was conducted on the Rhino Grasshopper platform \footnote{\url{https://www.grasshopper3d.com/}}, establishing an automatic optimization program for urban spatial design based on urban climate maps. Another method based on neural networks and genetic algorithms for landscape ecological construction and spatial pattern optimization was proposed to improve the efficiency of design optimization \cite{ban2022landscape}. This method utilizes computer vision information processing technology to analyze and extract the characteristics of the landscape ecological spatial environment, establishes a visual reconstruction model for landscape ecological construction and spatial pattern optimization design images, and enhances its ability to recognize and reconstruct the artistic features of the landscape ecological spatial environment. The results show that this algorithm not only reduces the number of iterations and runtime but also significantly improves the efficiency of landscape spatial optimization design. It has important practical guidance significance for guiding the construction of landscape ecological security patterns in various regions and promoting the process of green urbanization. For iterative updating of design solutions, the FloorplananGAN \cite{luo2022floorplangan} generates floor plans in vector format and uses CNN to visually differentiate the generated floor plans in a grid format. The native vector format generated by this algorithm ensures accuracy in editing, so the generated samples are represented by the coordinate points of room boundary boxes and can be exported in commonly used vector graphic formats such as DWG, SVG, and other formats. It can be integrated into CAD programs. FloorplanGAN provides an effective alternative to functional arrangement and layout optimization in the manual design process. The work from task list to preliminary layout can be completed automatically, thus accelerating the architectural design process and promoting iterative updates of design solutions.

\textbf{3) Multi-objective design optimization}. AIGC can apply multi-objective optimization algorithms \cite{deb2001multi} to help designers find the best design solutions while considering multiple objectives and constraints. By analyzing and optimizing different design variables and objective functions, AIGC can generate multiple optimized solutions for designers to choose from. AIGC can apply multi-objective optimization algorithms to balance and trade-off multiple objectives and constraints involved in design tasks, such as aesthetics, sustainability, cost, functionality, etc. Through AIGC-based parameterized design optimization, designers can quickly generate multiple design alternatives and flexibly adjust and optimize them during the design process, leading to more efficient and innovative designs. This approach fully leverages the advantages of AI technology to enhance design efficiency and quality, opening up new development opportunities in landscape design. Urban insights, based on the core technology DBF engine\footnote{\url{https://www.digitalbluefoam.com/}}, is a web-based urban design and planning software. It aims at supporting a more sustainable and livable future through data-driven decision-making. It includes three main components: visualization, prediction, and assessment, covering urban-level analysis, community and plot analysis, massing generation, design generation, and optimization evaluation. The software uses multi-source data and DBF's proprietary KPI cross-scale dataset for analysis and evaluation, including environmental reports, land-use data, diversity indices, urban morphology, social heatmaps, wind speed assessment, solar modeling, and other multi-objective analysis and optimization. We can also utilize genetic algorithms for multi-objective optimization. In a study \cite{yao2021generative}, residential building layout data of the Shanghai area learned by the Pix2pix Network was first used to train the Octopus genetic algorithm tool in the Grasshopper platform, generating different design solutions. During the generation process, optimization of multiple metrics such as daylight duration, ventilation performance, building aesthetics, and energy-saving cost was achieved by adjusting different training sample sets and using the Pareto genetic algorithm \cite{horn1994niched}, while effectively controlling plot ratio, building density, and height.

\subsection{Plant Configuration and Simulation}

\textbf{1) Automated plant configuration and simulation}. \textbf{a) Plant database and recommendations}: AIGC can build a vast plant database and recommend suitable plant selections based on site characteristics and requirements. By analyzing plant growth requirements, stress tolerance, and aesthetic features, AIGC can generate personalized plant recommendation lists to assist designers in plant selection and configuration. AIGC can enhance plant modeling and visualization capabilities, such as by using machine learning techniques and Terrestrial Laser Scanning (TLS) \cite{wu2021application}, which has demonstrated effectiveness in replacing manual sampling of various common indicators in forest ecosystems, it becomes feasible to model and visualize the canopy coverage and biomass of understory vegetation. A study \cite{anderson2018estimating}. employed a decision tree-based model developed through a random forest algorithm to predict biomass and canopy coverage across various vegetation categories. This was achieved using high-definition TLS scans of 26 one-ha plots in desert grasslands and big sagebrush shrublands in southwest Idaho, USA. This study demonstrated the ability of machine learning to meet current design and management needs by enhancing the time-consuming processes of creating vegetation inventories, modeling vegetation structural characteristics, environmental modeling studies, and the ability to train on a wider range of data sets collected from from aerial and satellite sources. Another study employed image interpretation technology, eCognition \cite{zhang2018practice,zhang2020AnRS}, to extract vegetation information and utilized a fuzzy classification algorithm supported by a decision expert system for intelligent image analysis. The study proposed an object-oriented classification technique for extracting information on typical vegetation landscape elements, improving the automatic identification accuracy of spatial image data and providing a novel quantitative analysis technique for the study of landscape vegetation features. \textbf{b) Plant growth simulation}: AIGC can simulate the growth process and evolution of plants, predicting growth rates, crown morphology, and landscape effects. Through the analysis of plant growth simulations, plant configuration and layout can be optimized, providing more accurate plant selection recommendations. For example, the L-System construction method \cite{xu2006fuzzy} based on the language description of plant growth has been applied to plant growth modeling and simulation. AI techniques are used to construct relevant fuzzy sets and L-System production rules through fuzzy inference. The growth process of plants is simulated in conjunction with carbon dioxide concentration in the plant growth environment. There is research proposing an integrated system for 3D tree modeling and growth simulation \cite{tang2015integrated}. This system is based on ontological knowledge bases and AI techniques (RBP, CBP), and it extends the existing 3D tree modeling software called ParaTree to simulate 3D tree structure modeling and growth under different conditions. \textbf{c) Plant classification and configuration}: AIGC can generate different plant combinations and ratios by analyzing the morphological, color, and texture features of plants. Through simulating and optimizing the effects of different plant combinations, diverse and innovative plant configuration schemes can be provided. Research has recorded measurements of environmental factors (water intake, light, temperature, humidity) and self-factors (plant weight, height, etc.) through sensors. By learning and analyzing the data using machine learning (ML) models, the analysis of the interaction between landscape plants and the environment can be obtained \cite{fahlgren2015lights}. The research \cite{fuentes2017robust} uses CNN combined with real-time image detection of plants to detect pests and diseases in landscape plants, achieving preventive effects. Some other studies \cite{gao2018fusion,hamylton2020evaluating} use image processing algorithms to monitor and classify plant landscapes on a large scale, capable of handling complex natural scene images and accurately classifying field plants. These image-processing algorithms employ CNN and random forest algorithms to learn detection from field images captured by unmanned aerial vehicles (UAVs).

\textbf{2) Plant generation design and simulation}. AIGC can perform environmental simulations to predict and evaluate the performance of design schemes under different environmental conditions. AI can use a large amount of environmental data and design parameters for simulation and prediction, helping designers understand the effects of plant growth, hydrological cycles, energy utilization, etc., and guide designers in making more accurate plant design decisions and optimizing design schemes. AIGC can generate realistic landscape visualizations and animations to assist designers and stakeholders in better understanding and evaluating design schemes. By simulating elements such as lighting, materials, and vegetation, realistic landscape effects can be presented, improving the quality of visual representation. AIGC can create virtual reality (VR) or augmented reality (AR) experiences. Note that VR is defined as "a computer-simulated environment that allows users to interact in the environment as if they are in the real world" \cite{rangelova2018survey}. This allows designers and stakeholders to experience plant design schemes firsthand. Through the interactivity and immersion of virtual environments, AIGC can help users better understand and participate in the design process. In terms of plant generation design, StyleGAN was proposed to learn the features and styles of floral images and generate new floral images \cite{xiang2023generated}. StyleGAN is a variant of a GAN that can learn and generate high-resolution and diverse images \cite{karras2020analyzing}. Numerous flower images are collected as training data through public flower image databases or flower photos taken by yourself. Use the preprocessed flower image dataset to train the StyleGAN model. During the training process, the model will learn the characteristics and style of flower images, such as the shape, color, texture, etc. of corollas, stamens, receptacles, etc., and generate new flower images. By adjusting the input vector of the model, it can accurately control the generation of flower images with different characteristics and styles. This method provides new possibilities for areas such as artistic creation and landscape design and shows great potential and value.

\subsection{Construction Management and Optimization}

In the construction and management of LA, AIGC can be applied in multiple aspects to improve construction efficiency, optimize management processes, and enhance project sustainability. Here are some applications of AIGC in LA construction and management:

\textbf{1) Intelligent construction and management}. AIGC provides intelligent decision support during the construction process by analyzing and learning from a large amount of construction data. For example, it can intelligently schedule construction personnel and equipment based on the site conditions and progress, thereby improving construction efficiency and quality. One mature technology for intelligent construction and management in LA is landscape information modeling (LIM), which evolved from building information modeling (BIM). With the advantage of software like Revit, LIM enables high collaboration among professionals, precise quantity extraction, and digital simulation of construction. Revit, mainly used in architecture, can be customized to break through barriers in its application to LA. In pilot project models, comprehensive models covering various disciplines (landscape architecture, water and electricity, greening, and structures) have been validated. To reduce the learning curve for Revit and enable automatic conversion from 2D drawings to 3D models, there has been successful piloting of generating greening plant models with one click, including cost and plant information data, linking standard structure samples to landscape objects, parameterized LA components, and automatic arrangement of landscape water and electricity sprinklers. LIM is specifically designed for information modeling in LA and planning. By integrating GIS, CAD, and BIM technologies, LIM creates a comprehensive 3D model of the landscape, including terrain, vegetation, water features, and infrastructure. This model can be used to analyze and simulate various scenarios, such as environmental impacts, water flow, and visual effects. The concept of LIM is to provide a holistic view of the landscape, enabling landscape architects and planners to design and manage landscapes more sustainably and efficiently. LIM can help reduce the environmental impact of landscape development, improve water and resource management, and enhance the overall quality of landscapes. The development of LIM is driven by technological advancements and the increasing demand for sustainable and integrated landscape design. The lack of standardization in LIM data and software may make it difficult to share and exchange data between different tools and platforms. AI technology can provide automated data processing, analysis, and modeling methods. AI can enhance LIM in various ways, such as automated feature extraction, machine learning algorithms, real-time simulation, and visualization. By integrating AI with LIM, landscape architects and planners can create more intelligent, dynamic, and responsive landscapes, while improving the efficiency and accuracy of the design process and planning process.

\textbf{2) Optimization of construction processes}. AIGC can optimize and improve construction processes through data analysis and machine learning techniques. By continuously monitoring and analyzing construction data, AIGC can identify potential issues and areas for improvement in the construction process, propose solutions, and adjust construction strategies in real-time. For example, Shunyin Zhiku\footnote{\url{https://www.shunyinzhichuang.com}} is a landscape intelligence system that aims to provide intelligent solutions for landscape design and garden planning to improve efficiency, accuracy, and innovation. Designers can quickly generate innovative design solutions and make real-time adjustments and optimizations using the system. The system utilizes big data analysis techniques to analyze various factors in landscape design, including terrain, climate, and vegetation. Based on the results of data analysis, the system can predict the effects and impacts of different design solutions, helping designers make more scientific and accurate decisions. In addition to intelligent research in the design phase, LIM in the construction drawing design phase enables ultimate visualization and digitization in the dynamic process of landscape design and construction. The demand for real-time monitoring and intelligent feedback has led to the development of BIM \cite{volk2014building} towards the direction of digital twin (DT) \cite{jones2020characterising}, which facilitates instant bidirectional integration of information-physical systems and supports intelligent decision-making. It emerges from the development of BIM through four stages and integration with other technologies, involving the use of sensors, simulation, and AI at different levels. Based on BIM, DT can integrate information in both directions throughout the facility lifecycle \cite{deng2021bim}. Related studies have explored the automatic generation of water and electricity points in construction drawings. In cases where the project terrain is complex, the cost-optimal and widely covered water and electricity irrigation system is a typical combinatorial optimization problem in operations research. To address this problem, there was a study focused on the application of ChatGPT in construction project risk management (CPRM) \cite{nyqvist2024can}, specifically in the areas of risk identification, analysis, and control. By using a mixed-method approach, the study qualitatively and quantitatively compared the performance of 16 risk management experts from a Finnish construction company with ChatGPT through anonymous peer review. The research showed that ChatGPT demonstrated excellent capabilities in generating comprehensive risk management plans, with quantitative scores significantly surpassing the average level of human experts. This indicates the potential of AIGC in improving construction project risk management.

\section{Potential Challenges and Trends} \label{sec:trends}
\subsection{Potential Challenges}

AIGC possesses numerous advantages in LA design, including data processing and analysis, design innovation and inspiration, visualization, and communication. However, it also faces several challenges, such as data quality and reliability, design expertise and judgment, user needs and engagement, among others. Specifically, the challenges faced by AIGC in the field of LA include, but are not limited to, the following aspects:

\textbf{1) Data quality and reliability}. The effectiveness and accuracy of AIGC are limited by the data quality (e.g., authenticity, accuracy, objectivity, diversity, etc.). If the data is inaccurate, incomplete, or biased, the design solutions generated by AIGC may be affected. Therefore, it is necessary to ensure the data's accuracy, completeness, and consistency to enhance the accuracy and reliability of AIGC in the domain knowledge. High-quality data can significantly improve AI models' performance and generalization ability, reducing overfitting and underfitting phenomena. The field of LA is complex and knowledge-intensive, and if the plant database upon which the system relies contains erroneous or outdated information, it may recommend unsuitable plant species. The data sources and algorithms used by AIGC may also have biases or errors. Designers need to validate and review the data to ensure its objectivity, reliability, and applicability. Data cleaning and selection are currently core bottlenecks for AI algorithmic intervention. AIGC technology can be further utilized to precisely control the distribution and domain characteristics of datasets, avoiding the noise and biases present in real-world data and generating LA data that are difficult to obtain or lacking (thus enhancing diversity).

\textbf{2) Design expertise and judgment}. The use of AIGC may trigger discussions regarding the role and value of designers. Designers should clarify their roles in AIGC and leverage their professional knowledge, judgment, and creativity to ensure the quality and uniqueness of design solutions. For example, an AI system may generate many design solutions, but they may appear stylistically similar and lack personality. It cannot entirely substitute for the creative thinking and aesthetic concepts of human designers. AIGC still cannot fully replace human creativity and imagination. The designer's unique perspective, aesthetic taste, and creative thinking remain indispensable factors in the design process. For instance, a system may generate a realistic landscape image, but it may fail to convey the underlying design concepts and emotions. These aspects require the involvement of the designer's role. AI systems may not fully consider the differences in cultures and regions. For example, in an international project, a system may generate a plant that is popular in a specific region but may not be well-accepted or suitable for the local environment.

\textbf{3) Technical challenges and limitations}. AIGC requires more comprehensive and accurate acquisition and analysis of site data and environmental factors to provide more precise design support. Additionally, AIGC needs to better understand and meet user needs while providing more participation and feedback mechanisms to ensure that design solutions align with user expectations. The application of AIGC is restricted by the current limitations of technologies. Although progress has been made in creative tasks using AI, there are still limitations, such as the quality of image and sound synthesis and the accuracy of semantic understanding. Currently, AI is primarily used for generating images and cannot perform fine-grained modeling. However, LA design cannot solely rely on image representation, making precise adjustments challenging. Therefore, the solutions generated by the current AI technology cannot replace the final rendering of design solutions submitted to clients. Furthermore, developing smarter and more innovative algorithms and models to better simulate the human creative process and achieve bidirectional communication and collaboration between creativity and technology is another challenge.

\textbf{4) Site characteristics and sustainability}. Sustainable site design involves analyzing the existing conditions and circumstances of a site and implementing an overall layout plan. AIGC may not fully consider the site's characteristics and environmental factors. For example, a system might suggest planting a specific type of plant on a particular site without taking into account the climate and soil conditions of the area, resulting in the plants being unable to survive. The sustainability of landscape design encompasses ecological sustainability, economic sustainability, and social sustainability, all of which are interconnected. AI-generated designs may not adequately consider sustainability and ecological conservation factors. For instance, when designing water features, a system may generate large ponds or fountains without considering the sustainable use of water resources, economic costs, and the balance of ecosystems. Sustainability is a shared challenge for humanity and an issue that AIGC needs to pay special attention to.

\textbf{5) User needs and engagement}. AI-generated design solutions may not fully meet the individualized needs of various users. Designers should refine and customize the designs generated by AIGC to address specific user requirements. Each user has unique aesthetic preferences and personal preferences, while an AI system may only provide generalized design solutions. For example, a system may generate a set of design proposals that do not meet the specific requirements of users for privacy and comfort. The use of AIGC in the design process may reduce user engagement and involvement. Designers should facilitate user participation and feedback through other means such as workshops and discussions to ensure that design solutions truly meet their expectations and needs. Developing personalized and customized AI tools and platforms based on different design fields and creative needs is essential to meet the specific requirements of creators.

\textbf{6) Technology and creativity win-win}. AIGC faces the dual challenges of technology and creativity in the creative field, requiring a balance between technological advancements and artistic design. In terms of artistic and design creativity, humans possess unique subjective aesthetics and creativity. While AI can imitate and generate art or design works, maintaining the artist's unique style and creative thinking while staying technologically advanced is a key challenge. AIGC may lack the consistent design style and aesthetic of a design team. The unique understanding and diversity in design are crucial advantages in the industry. Therefore, technology must collaborate closely with design to explore the seamless integration of technology and art. In terms of technology, interdisciplinary collaboration between technical experts and designers is crucial for jointly exploring methods to combine AI technology with design creativity, resulting in more unique and technically advanced works. Respecting originality and intellectual property rights while being vigilant about potential copyright infringement and piracy issues brought about by AI is important. Through technological innovation, interdisciplinary collaboration, and considerations of social ethics, a win-win situation can be achieved for AI in the creative field, fostering the fusion of art and technology and providing greater innovative momentum for human society.

\textbf{7) Ethics and social impact}. AIGC relies heavily on data during the design process, and the quality and reliability of the data are crucial for accurate design outcomes. AIGC needs to address issues related to data collection, cleaning, and verification to ensure the accuracy and reliability of design decisions. For example, if a system relies on copyrighted design works for learning and generation, it may raise some issues of intellectual property infringement. AIGC faces regulatory and legal challenges. An example of an AI art generator copyright dispute: a court in Northern California ruled in favor of Midjourney and Stability AI in a copyright infringement case, but the controversy continues. AI generators are trained on human work, which is often unauthorized. The application of AIGC also involves ethical and social impact issues. For instance, if a system is widely used in landscape design, resulting in numerous automated design solutions, it may reduce employment opportunities for traditional designers and have adverse effects on the industry.

These challenges should be considered comprehensively and balanced in the applications of AIGC. Designers should fully understand the potential and limitations of AIGC, utilize it flexibly, and combine it with their professional knowledge and judgment. By integrating AI technology with the creative thinking and expertise of human designers, better landscape design outcomes can be achieved. So far, we have provided a summary of the potential, future trends, and prospects of AIGC in landscape design, as well as a discussion and outlook on the technological and ethical challenges it faces.

\subsection{Future Trends}

The future development trends and prospects of AIGC in landscape design are mainly reflected in the following aspects, including but not limited to:

\textbf{1) Interdisciplinary collaboration driving technological advancements}. LA encompasses various disciplines such as ecology, agronomy, forestry, geography, architecture, urban planning, sociology, art, psychology, and more. The integration of AI in landscape design expands its scope to include computer science, geographic information systems, aerial remote sensing, satellite positioning, and other interdisciplinary fields. The synthesis, systematization, and integration of interdisciplinary knowledge facilitate the merging of innovative resources, interdisciplinary collaboration, and multi-team cooperation. Government departments should establish policies and measures to support interdisciplinary research, providing a favorable policy environment and institutional guarantees to promote the continuous development of AI technologies. This will further enhance AIGC's design capabilities, optimization of design solutions, data analysis, and interactive capabilities.

\textbf{2) Comprehensive technological applications covering the LA field}. AIGC is becoming increasingly powerful, enabling intelligent generation of text, images, audio, video, and more. Therefore, AIGC will be applied to a wider range of landscape design, construction, and management tasks, including urban park green spaces, urban square designs, residential area landscapes, tourist attraction designs, historical and cultural heritage preservation, comprehensive landscape engineering and management, as well as public services (e.g., public participation and visitor guide services). For example, a digital landscape management platform can be established to integrate various technological methods and data resources, enabling the effective management and monitoring of landscape projects. In the future, there will be increased development efforts in AIGC, leading to more widespread applications in various areas and facilitating comprehensive technological advancements.

\textbf{3) Data support, algorithm optimization, and enhanced collaboration with designers}. The application of AI technology in LA will involve stronger data support and algorithm optimization to improve the reliability and accuracy of AI technology. Collecting and organizing extensive landscape design-related data, including topographic data, climate data, and vegetation distribution data, provides essential support for AIGC. Additionally, optimizing AIGC algorithms enhances their applicability and efficiency in landscape design. In LA, the application of AI technology necessitates close collaboration with designers. Establishing a mechanism for close cooperation between landscape designers and AI experts, jointly exploring the challenges and needs of landscape design, and adjusting AIGC algorithms based on real-world scenarios can fully harness designers' creativity and AI's technological advantages, ultimately improving the quality and efficiency of designs.

\textbf{4) Seamless integration with AI technology, expanding research perspectives and scales}. AIGC will integrate with other technologies such as data mining, big data, VR, and AR to provide designers and users with more immersive design and interactive experiences. Through the fusion of multisource data, including geographic information system data, remote sensing data, and socioeconomic data, landscape research can benefit from multi-perspective and multi-scale research data, spanning urban plans to urban facades and encompassing both macro and micro levels. AI technology can be used for data integration and analysis, expanding research perspectives. Deep learning models can efficiently model and predict complex landscape systems. More intelligent design tools, such as large language models (LLMs) \cite{gan2023model,lai2024large,zheng2024large} will be developed, combining AI and human-computer interaction technologies, enabling designers to flexibly explore designs at different scales.

\textbf{5) New ethical standards, laws, and regulations will continue to be improved}. With the continuous advancement of AIGC applications, ethical, privacy, and security issues should be considered in specific designs. In particular, existing laws and regulations are no longer able to constrain and protect against potential infringement issues arising from generative design. Relevant ethical standards will be further improved to ensure the legality and security of designs. For example, clarifying the responsibilities and compensation liabilities of AI systems, establishing corresponding legal liability frameworks, and implementing ethical review and regulatory mechanisms for AI technology. By formulating new ethical standards, laws, and regulations, the application and development of AI technology can be effectively regulated, safeguarding the rights and safety of the general public and promoting the healthy development of AI technology.

\section{Conclusions}  \label{sec:conclusions}

The application of AIGC in landscape design has broad prospects and potential. It provides support in various aspects, such as rapid design exploration, data-driven decision support, intelligent site assessment, personalized design solutions, visualization and interactive experiences, design optimization and efficiency improvement, sustainability, and ecological preservation. However, AIGC still faces challenges in landscape design, such as expressing creativity and uniqueness, considering site characteristics and environmental factors, and satisfying user needs and engagement. Through in-depth research and exploration, we can better explore the potential of AIGC in landscape design and address corresponding challenges to promote its application and development in the field. In this paper, we first comprehensively summarize the current status of LA, including its complexity and diversity, as well as its challenges and limitations. We then specifically discuss AIGC's application potential and key technologies in LA, including site research and analysis, design concepts and solution generation, parametric design optimization, plant configuration and simulation, construction management, and optimization support. Finally, we highlight the relevant issues and challenges in LA, the technological barriers, and the development trends and prospects, which require interdisciplinary integration, technological and creative support, sound laws and regulations, and the indispensable subjective aesthetics of humans. Behind the landscape technologies of AIGC lie aesthetics, psychology, and keen observations of life and society.

\printcredits

\bibliographystyle{cas-model2-names}

\bibliography{aigcla.bib}

\end{document}